\documentclass[11pt,draftcls,onecolumn,journal]{IEEEtran}
\usepackage{amsmath,amsfonts}
\usepackage{algorithmic}
\usepackage{algorithm}
\usepackage{array}
\usepackage[caption=false,font=normalsize,labelfont=sf,textfont=sf]{subfig}
\usepackage{textcomp}
\usepackage{stfloats}
\usepackage{url}
\usepackage{verbatim}
\usepackage{graphicx}
\usepackage{cite}
\usepackage{hyperref}
\usepackage{booktabs}
\usepackage{xcolor}
\begin{document}

\title{Image Modeling:
From Handcrafted Priors to Foundation Image Models}

\author{A. Murat Tekalp,~\IEEEmembership{Life Fellow,~IEEE}
\thanks{The author is with the Department of Electrical and Electronics Engineering, Koç University, Istanbul, Turkey}
\thanks{Manuscript received April 5, 2026, revised June 26, 2026, revised August 12, 2026, revised again October 10, 2026.}
}

\markboth{accepted to IEEE Signal Processing Magazine, October~2026}%
{\MakeLowercase{\textit{et al.}}: Author Guidelines for Special Issue Articles of IEEE SPM}

\maketitle

\vspace{-52pt}
\IEEEPARstart{C}onsidering a $1000 \times 1000$ pixel grid, we have a $1,000,000$ dimensional space. If we assign random colors to each pixel, the resulting array almost surely fails to resemble a natural image.
Natural~images exist on a very thin manifold within that high-dimensional space. 
Image modeling refers to the~mathematical description of this manifold. Such models provide priors that act as constraints~to compute point estimates for inverse problems, or act as generative models that can be sampled to produce new images. 
Recent developments have led to the unification of both tasks under foundation image models.

\begin{figure}[h]
\begin{center}
\includegraphics[width=0.99\textwidth]{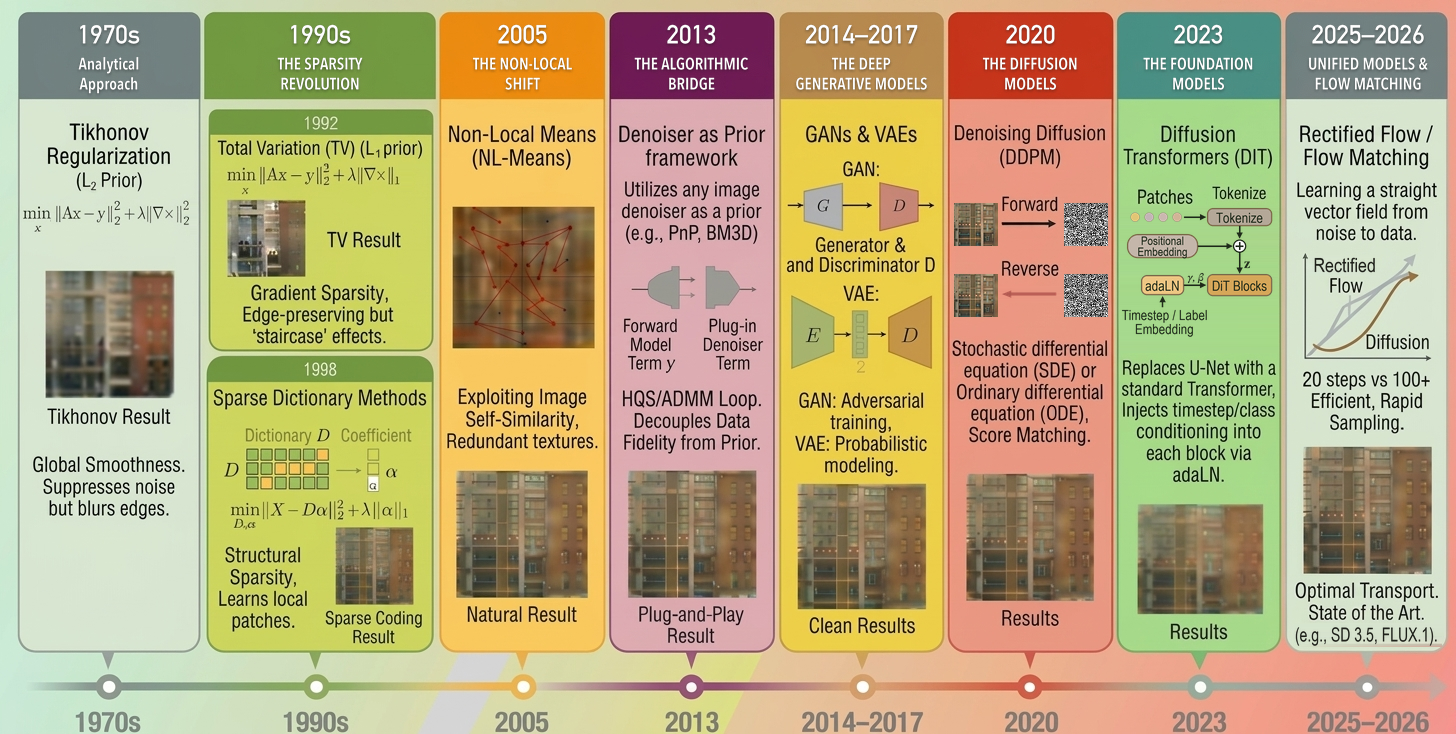} \vspace{-18pt}
\caption{The timeline of evolution of image modeling from hand-crafted priors to foundation image models. Illustration is generated using Google Gemini controlled by author prompts. The layout is designed and the content is verified by the author.}
\vspace{-24pt}
\end{center}
\label{fig:sonn-sr}
\end{figure}

This paper unifies six decades of image modeling—evolving from handcrafted mathematical tractable priors to solve inverse problems in imaging to data-driven deep foundation image models for both image generation and solving inverse problems—under a common Bayesian framework that formulates image modeling as an  explicit or implicit probabilistic characterization of a data manifold. Every image restoration or generative task fundamentally seeks to answer the same question: What makes an image look like a real image? 
In the Bayesian framework, this question is encapsulated by the~prior probability density $p({\bf x})$, or its more optimization-friendly counterpart, the~log-prior $\log p({\bf x})$. 
While early works in solving inverse problems, such as Tikhonov, Total Variation (TV), and Non-local priors relied on explicit handcrafted definitions of log prior to model local smoothness, modern techniques, such as Plug-and-Play~(PnP) priors and Denoising Diffusion models, are rooted in algorithmic uncoupling and implicit manifold learning that apply to both inverse problems and conditional image generation. By grounding imaging inverse problems and conditional image generation under a single mathematical engine: the~gradient of the~log-prior, or the Stein Score Function $\nabla_{\bf x} \log p({\bf x})$, we bridge the gap between signal processing and machine learning communities, turning what seems like parallel developments into a single, evolution of image modeling from handcrafted analytical models to models learned from big data, demonstrating that the~pursuit of the 'denoising score' is the common thread connecting variational optimization (in MAP estimation) to posterior sampling (in modern foundation image models).

\section*{Image Modeling for Solving Inverse Problems vs. Conditional Image Generation}
Both solving imaging inverse problems and conditional image generation (e.g., text-to-image (T2I)) can be formulated under a shared probabilistic framework. While both tasks can be written using the~identical conditional notation $p({\bf x}|{\bf y})$, they differ fundamentally in the nature of the conditioning variable~${\bf y}$, such as presence of an explicit physical forward model and the uncertainty of the resulting distribution. They also differ in their solution strategies: While inverse problems are fundamentally ill-posed Bayesian inversion tasks that require navigating a forward model and can be solved via either point optimization (MAP estimation) or posterior sampling; conditional image generation models the broader conditional distribution $p({\bf x}|{\bf y})$ directly, which can be evaluated via regression or conditional sampling.

\subsection*{Imaging Inverse Problems}
We observe a degraded, noisy measurement ${\bf y} \in \mathbb{R}^m$ of the clean image ${\bf x} \in \mathbb{R}^n$, 
$${\bf y} = \mathbf{A}{\bf x} + \mathbf{n}$$ 
where $\mathbf{A}$ is called a forward operator and $\mathbf{n}$ denotes observation noise. 

According to Bayes' rule, the~posterior distribution is:
$$p(\mathbf{x}|\mathbf{y}) = \frac{p(\mathbf{y}|\mathbf{x})p(\mathbf{x})}{p(\mathbf{y})} \propto p(\mathbf{y}|\mathbf{x})p(\mathbf{x})$$

In the Bayesian formulation of inverse problems, there are two different major solution paradigms: Maximum A Posteriori (MAP) estimation and generative posterior sampling. Both approaches operate on the exact same posterior distribution $p(\mathbf{x}|\mathbf{y})$, but MAP estimation is an optimization problem, whereas posterior sampling is a simulation problem.
\vspace{-20pt}

\subsubsection*{Point Estimation}
MAP condenses the posterior distribution to a single point estimate—the mode, which is the single most likely clean image $\hat{\mathbf{x}}$ that fits both the data and the prior, given by
\begin{eqnarray}
\hat{\mathbf{x}}_{\text{MAP}} &=& \arg\max_\mathbf{x} \log p(\mathbf{x}|\mathbf{y}) = \arg\max_\mathbf{x} \left[ \log p(\mathbf{y}|\mathbf{x}) + \log p(\mathbf{x}) \right] \nonumber \\
&=&\arg \min_\mathbf{x} \{ - \log p(\mathbf{y}|\mathbf{x}) - \log p(\mathbf{x}) \}=\arg \min_\mathbf{x} \{ \underbrace{D(\mathbf{x},\mathbf{y})}_{\text{Data Fidelity Term}} + \underbrace{\mathcal{R}(\mathbf{x})}_{\text{Prior/Energy Term}} \}
\end{eqnarray}
where $D(\mathbf{x},\mathbf{y})$ measures fidelity to the observed data (e.g., $D(\mathbf{x};\mathbf{y})=\frac{1}{2}\|\mathbf{Ax}-\mathbf{y}\|_2^2$ for Gaussian noise), which is defined by physics and the prior \vspace{-6pt}
$$\mathcal{R}(\mathbf{x}) = - \log p(\mathbf{x})$$ is the core of this presentation. It defines 'what a clean image looks like.' This energy function is minimized when $\mathbf{x}$ lies on the data manifold.
In signal processing, this is solved using optimization algorithms (e.g., ADMM, proximal gradient descent) to find the peak of the probability density landscape.

Point estimation minimizes distortion at the expense of lower perceptual quality. 
When faced with an ambiguous null-space (e.g., missing frequencies in super-resolution), for example, minimum mean square error (MMSE) estimation averages all possible plausible solutions to minimize pixel-to-pixel mean-squared error (MSE). In image restoration, this typically causes loss of fine, high-frequency textures.

\subsubsection*{Posterior sampling}
Posterior sampling draws samples $\mathbf{x}^{(i)}$ directly from the posterior distribution \vspace{-6pt}
$$\mathbf{x}^{(i)} \sim p(\mathbf{x}|\mathbf{y}) \vspace{-4pt}$$
This is achieved using score-based diffusion guidance, flow matching, or Langevin dynamics to navigate from noise to image samples.
To sample from this posterior using score-based generative models or Langevin dynamics, we compute the gradient of the log-posterior (the posterior~score): \vspace{-4pt}
$$\nabla_\mathbf{x} \log p(\mathbf{x}|\mathbf{y}) = \underbrace{\nabla_\mathbf{x} \log p(\mathbf{y}|\mathbf{x})}_{\text{Physical Likelihood Score}} + \underbrace{\nabla_\mathbf{x} \log p(\mathbf{x})}_{\text{Prior Score}}$$


Posterior Sampling maximizes perceptual quality at the expense of higher distortion since it does not average possibilities.
The choice between MAP and posterior sampling directly depends on the~desired visual characteristics of the reconstruction. The tradeoff between fidelity vs. perceptual quality is formalized by the perception-distortion tradeoff proposed by Blau \& Michaeli \cite{bla18}.
\vspace{-10pt}

\subsection*{Conditional Image Generation: Text-to-Image}
Given a semantic text prompt or conditioning context $c$, the goal of conditional generation is to sample from the conditional distribution $p_t(\mathbf{x}_t|c)$ along the generative trajectory. Applying Bayes' rule yields: \vspace{-5pt}
$$\nabla_{\mathbf{x}_t} \log p_t(\mathbf{x}_t|c) = \underbrace{\nabla_{\mathbf{x}_t} \log p_t(c|\mathbf{x}_t)}_{\text{Implicit Classifier Score}} + \underbrace{\nabla_{\mathbf{x}_t} \log p_t(\mathbf{x}_t)}_{\text{Unconditional Score}} \vspace{-3pt}$$
where $t \in [0, T]$ denotes the noise level index (diffusion time step) along a generative trajectory.

In practice, computing the classifier score $\nabla_{\mathbf{x}_t} \log p_t(c|\mathbf{x}_t)$ directly via an external, noise-aware image classifier is notoriously unstable and prone to adversarial artifacts. To bypass this, modern foundation models employ Classifier-Free Guidance (CFG). By jointly training a single network to estimate both the~conditional score and the unconditional score (via random condition dropping), the generation process can extrapolate between them during sampling using a guidance scale $w$:  
$$\tilde{\mathbf{s}}_t(\mathbf{x}_t, c) = (1+w)\nabla_{\mathbf{x}_t} \log p_t(\mathbf{x}_t|c) - w \nabla_{\mathbf{x}_t} \log p_t(\mathbf{x}_t)$$
This algebraic formulation allows the model to amplify the conditioning signal and trade off sample diversity for visual fidelity without ever explicitly evaluating a classifier.
\vspace{-10pt}

\subsection*{Distinctions between Imaging Inverse Problems and Conditional Image Generation}
Even though both reduce to evaluating a conditional score function, they exhibit different behaviors: In inverse problems, the likelihood term $p(\mathbf{y}|\mathbf{x})$ is defined by physics and noise statistics. If we know the~forward operator $\mathbf{A}$ and assume Gaussian noise, $p(\mathbf{y}|\mathbf{x}) = \mathcal{N}(\mathbf{y}; \mathbf{Ax}, \sigma^2 \mathbf{I})$, the gradient 
$$\nabla_\mathbf{x} \log p(\mathbf{y}|\mathbf{x}) = \frac{1}{\sigma^2}\mathbf{A}^T(\mathbf{y} - \mathbf{Ax})$$ 
can be computed exactly by linear algebra.
In T2I, the term $p(c|\mathbf{x})$ describes the probability that a human would write the text description~$c$ given the image $\mathbf{x}$. This is a semantic mapping that cannot be expressed as a closed-form equation. It must be learned from data pairs using alignment networks (e.g., CLIP \cite{rad21}).

\vspace{-10pt}

\subsection*{Text-Guided Image Restoration: Combining Inverse Problems and T2I}
Understanding that both problems are conditional sampling tasks allows us to combine them into a unified Text-Guided Image Restoration framework: conditional sampling from $p_t(\mathbf{x}_t|\mathbf{y}, c)$ along a generative trajectory.

When leveraging a Foundation Image Model to solve an inverse problem, we assume the physical measurement $\mathbf{y}$ and the semantic context $c$ are conditionally independent given the true image state. This allows us to decompose the total time-dependent score function into a linear combination of vector fields:
$$\nabla_{\mathbf{x}_t} \log p_t(\mathbf{x}_t|\mathbf{y}, c) = \underbrace{\nabla_{\mathbf{x}_t} \log p_t(\mathbf{y}|\mathbf{x}_t)}_{\text{Enforced Physics Constraints}} + \underbrace{\nabla_{\mathbf{x}_t} \log p_t(\mathbf{x}_t|c)}_{\text{Enforced Semantic Context}}$$
Because the physical likelihood $p(\mathbf{y}|\mathbf{x}_0)$ is defined in the clean image domain, the gradient of the~data-fidelity term (the likelihood) cannot be evaluated directly on the noisy latent $\mathbf{x}_t$. Instead, implementations, such as Diffusion Posterior Sampling, utilize the foundation model to estimate the clean image $\hat{\mathbf{x}}_0(\mathbf{x}_t)$ at each step, passing the physical residual back through the network via the chain rule to enforce consistency.

In summary, imaging inverse problems and text-to-image generation leverage complementary conditioning terms grounded on the exact same probabilistic framework to navigate the natural image manifold. 
\vspace{-10pt}

\section*{Traditional Image Priors: Hand-Crafted Models}
\label{traditional}

Before deep learning, researchers observed that natural images possess 
deterministic or statistical regularity properties, which allow for "regularization" in inverse problems in imaging. They expressed various handcrafted log priors, $\log p(\mathbf{x})$ explicitly to model regularity properties of natural images.
\vspace{-4pt}

\subsection*{Tikhonov Regularization: $L_2$ Smoothness Prior and Homogeneous Gaussian Random Fields}
Early image modeling, in the pre-1980 era, assumed image intensity varies smoothly over the spatial coordinates. 
These global smoothness constraints came in the form of deterministic or stochastic models.

Tikhonov regularization \cite{tik77}, also referred to as ridge regression in statistics, is the~classic deterministic approach to stabilizing ill-posed inverse problems by a variational optimization formulation, where the~regularizer $R(\mathbf{x})$ is defined as the squared ($L_2$) norm of the image gradient, \vspace{-2pt}
\begin{equation}  
R(\mathbf{x}) = \lambda \|\mathbf{\nabla x}\|_2^2
\end{equation}
Tikhonov regularization is mathematically equivalent to modeling images as a  realization of a homogeneous (stationary) constant-mean Gaussian random process, with the prior \vspace{-2pt}
\begin{equation}  
p(\mathbf{x}) \propto \exp{(-\lambda \|\mathbf{\nabla x}\|_2^2) }  \vspace{-2pt}
\end{equation}
which makes the score linear: $\nabla_\mathbf{x} \log p(\mathbf{x}) = -2\lambda \mathbf{\nabla^2 x}$, where $\nabla^2=\nabla \cdot \nabla$ denotes the Laplacian.

While mathematically elegant and computationally efficient due to linearity of its score—leading to closed-form solutions 
or simple iterative gradients—this global smoothness prior results in "oversmoothing" of sharp edges because the~$L_2$ penalty treats all variations with equal weight regardless of their structural significance.
The stationary Gaussian prior interpretation, which allows complete description of an image by its power spectral density, treating the image as a uniform texture, provided the theoretical foundation for early statistical image processing, such as Wiener filtering in the~frequency domain. However, it fails to account for the non-stationary nature of real-world images. Consequently, stationary models cannot adapt to abrupt changes in local image statistics and produce ringing artifacts near edges.

\subsection*{Non-stationary Mean Non-stationary Variance Gaussian Model}
The non-stationary-mean non-stationary-variance (NMNV) Gaussian model \cite{lee80,kuan85} is a stochastic framework that characterizes the local statistics of natural images. 
It is considered a cornerstone in the~transition from global (stationary) filters based on global smoothness priors to the spatially adaptive "edge-directed" filters in image processing.
Unlike earlier "stationary" models, the NMNV model allows the mean and variance to vary spatially to better capture local structures. 
The~image $x(i, j)$ is modeled~as:
\begin{equation}
x(i, j) = \mu_x(i, j) + \sigma_x(i, j)  w(i, j)
\end{equation}
where $\mu_x(i, j)$ is the local average intensity (non-stationary mean), capturing low-frequency structure, $\sigma_x(i, j)$ (non-stationary variance) capturing local activity, e.g., edges, textures, and high-frequency details, and $w(i, j)$ is a zero-mean, unit-variance white Gaussian noise. It states that if we subtract the~local mean $\mu_x(i, j)$ from the image $x(i, j)$, the difference can be modeled as spatially-varying scaled white noise.

This model has been the basis for the Kuan filter \cite{kuan85}, which is a minimum mean square error (MMSE) estimator for image restoration based on local image statistics. By allowing $\mu$ and $\sigma$ to change at every pixel, 
the model applies stronger smoothing to suppress noise in low-variance regions, while preserving the signal in high-variance regions (edges and texture) to avoid blurring critical details.
\vspace{-6pt}

\subsection*{Markov Random Fields} 
Markov Random Field (MRF) is a probabilistic framework to model the spatial context of an image~by assuming the state of a pixel depends only on its spatial neighborhood. The~application of MRFs to image modeling was revolutionized by the Hammersley-Clifford theorem that establishes equivalence between MRFs and Gibbs distributions. It allowed expressing the probability of an image $\mathbf{x}$ as a Gibbs measure:
$$p(\mathbf{x}) = \frac{1}{Z} \exp\left( - \sum_{c \in \mathcal{C}} V_c(\mathbf{x}) \right)$$
where $Z$ is the partition function, $\mathcal{C}$ denotes the set of local cliques (e.g., adjacent pixel pairs), and $V_c(\cdot)$~are the handcrafted potential functions that penalize local variations.
The seminal work of Geman and Geman~\cite{gem84} demonstrated the power of MRFs in image restoration and segmentation tasks.
However, MRFs rely on handcrafted local clique potentials to model piece-wise smoothness while preserving boundaries and struggle to capture the long-range  dependencies in high-resolution natural images. 
\vspace{-6pt}

\subsection*{Local Smoothness: Total Variation (TV) and Generalizations}
The~Total Variation (TV) prior assumes images consist of piecewise-constant areas of intensity separated by sharp~edges. Unlike Tikhonov regularization, which penalizes the $L_2$~norm of image derivatives, the~isotropic TV prior \cite{rud92} penalizes the spatial $L_1$ norm of Euclidean gradient magnitudes at each pixel (also referred to as un-squared mixed $\ell_{1,2}$ norm):
\begin{equation} 
R(x)=\int_\Omega \| \nabla x(u) \|_2 \, du
\end{equation}
or in discrete notation for a vector $\mathbf{x}$ as
\begin{equation}
R(\mathbf{x}) = \sum_{i,j} \sqrt{(x_{i+1,j} - x_{i,j})^2 + (x_{i,j+1} - x_{i,j})^2}
\end{equation}
This is equivalent to assuming an isotropic bivariate Laplace distribution on image gradients:
$$p(\mathbf{x}) \propto \exp(-\lambda R(\mathbf{x}))$$ 
Because the Euclidean gradient magnitude $\sqrt{g_1^2 + g_2^2}$ has a non-differentiable conical point at the origin, the Stein score function does not formally exist at zero. To mitigate this, the original ROF paper \cite{rud92} replaced the gradient magnitude with a smoothed approximation, now known as the Charbonnier penalty:
\begin{equation}
R_\varepsilon(\mathbf{x}) = \sum_{i,j} \sqrt{(x_{i+1,j} - x_{i,j})^2 + (x_{i,j+1} - x_{i,j})^2 + \varepsilon^2}
\end{equation}
where $0 < \varepsilon \ll 1$. Alternatively, it is possible to use the piecewise-smooth Huber loss.
Note that proximal algorithms (like Chambolle's Projection Algorithm or the Split Bregman/ADMM methods) to handle non-differentiable norms had not been invented yet when the ROF paper was published in 1992.

The TV prior penalizes the "total amount" of change in pixel intensities. Because of its spatial $L_1$ structure, it finds it "cheaper" to have one large jump (an edge) than many small random variations (noise). Hence, it flattens noise into smooth regions while leaving important edges sharp and intact.
However, modeling images as piecewise-constant blocks with sharp transitions generates artifacts known as the~"staircase effect," where regions with soft linear intensity variations appear as step-like stairs.

To overcome the limitations of the ROF model, several generalized TV frameworks were introduced. These include $L_1$-TV for impulsive noise, vectorial TV for multi-channel color imaging, non-local TV for texture preservation, and Total Generalized Variation (TGV) for mitigating staircasing artifacts. For a comprehensive review of these architectural variations and their numerical solvers, we refer the reader to Chambolle et al. \cite{cham10} and the space-variant TV under a Bayesian framework by Pragliola et al. \cite{pra23}.
\vspace{-6pt}

\subsection*{Sparse Representations and Dictionary Learning Frameworks}

While natural images are dense in pixel space, they are "sparse" in some linear transform space. The~sparse representation paradigm models an image (or patches) as a linear combination of a small number of atoms, chosen from an overcomplete dictionary matrix. 
Given a fixed dictionary $\mathbf{D} \in \mathbb{R}^{m \times n}$ (where $m \ll n$), sparse coding maps an observed signal $\mathbf{y} \in \mathbb{R}^m$ into a latent coefficient vector $\boldsymbol{\alpha} \in \mathbb{R}^n$ via $\mathbf{y} = \mathbf{D}\boldsymbol{\alpha}$. The "sparse" part of the definition requires that most entries in $\mathbf{\alpha}$ are zero. There exist different mathematical formulations:

1. The Penalized Form  \cite{che98}: 
$$\min_{\boldsymbol{\alpha}} \frac{1}{2} \|\mathbf{y} - \mathbf{D}\boldsymbol{\alpha}\|_2^2 + \lambda \|\boldsymbol{\alpha}\|_p, ~~~~~~p\in\{0,1\}$$ \vspace{-2pt}
where $p$ defines the nature of sparsity constraint, and $\lambda$ balances reconstruction fidelity~vs.~sparsity.

2. The Sparsity-Constrained Form: \vspace{-2pt}
$$\min_{\boldsymbol{\alpha}} \|\mathbf{y} - \mathbf{D}\boldsymbol{\alpha}\|_2^2 \quad \text{subject to} \quad \|\boldsymbol{\alpha}\|_p \le T_0$$
where $T_0$ is a strict, predefined budget of allowed non-zero coefficients. This form is natively solved via greedy pursuit metrics like Orthogonal Matching Pursuit (OMP) \cite{pat93}.

3. The Error-Constrained Form (Basis Pursuit Denoising) \cite{che98}: \vspace{-2pt}
$$\min_{\boldsymbol{\alpha}} \|\boldsymbol{\alpha}\|_p \quad \text{subject to} \quad \|\mathbf{y} - \mathbf{D}\boldsymbol{\alpha}\|_2^2 \le \epsilon$$
where $\epsilon$ represents a bounded tolerance for noise or representation error.

If $p=1$, the objective functions and constraints are convex and these three forms are weakly equivalent, meaning they map out the same Pareto frontier.
If $p=0$, the $L_0$ "norm" counts non-zero coefficients, making the optimization problem non-convex and NP-hard, and the equivalence breaks down.

These optimization formulations can be generalized through a hierarchical Bayesian framework. Let's treat the latent coefficients $\mathbf{\alpha}$ as a random variable and the modeling error as additive noise $n$: $\mathbf{y} = \mathbf{D\alpha} + \mathbf{n}$. Under the MAP estimation objective, finding the optimal representation corresponds to minimizing:
$$\hat{\boldsymbol{\alpha}}_{\text{MAP}} = \arg\max_{\boldsymbol{\alpha}} \log p(\boldsymbol{\alpha}|\mathbf{y}) = \arg\min_{\boldsymbol{\alpha}} \Big[ -\log p(\mathbf{y}|\boldsymbol{\alpha}) - \log p(\boldsymbol{\alpha}) \Big]$$
Assuming an independent and identically distributed (i.i.d.) zero-mean Gaussian noise $\mathbf{n} \sim \mathcal{N}(0, \sigma^2 \mathbf{I})$, the conditional likelihood function takes the exponential form $p(\mathbf{y}|\boldsymbol{\alpha}) \propto \exp\left(-\frac{\|\mathbf{y} - \mathbf{D}\boldsymbol{\alpha}\|_2^2}{2\sigma^2}\right)$. Its negative logarithm generates the standard $L_2$ data fidelity term:
$$-\log p(\mathbf{y}|\boldsymbol{\alpha}) = \frac{1}{2\sigma^2} \|\mathbf{y} - \mathbf{D}\boldsymbol{\alpha}\|_2^2$$
Assuming that $\alpha_i$ are i.i.d. and follow a heavy-tailed Laplace distribution $p(\alpha_i) = \frac{1}{2b}\exp\left(-\frac{|\alpha_i|}{b}\right)$, the~log-prior yields the convex $L_1$-norm regularization seen in the LASSO framework:
$$-\log p(\boldsymbol{\alpha}) = \frac{1}{b} \|\boldsymbol{\alpha}\|_1 + C$$
\vspace{-14pt}

\subsubsection*{Dictionary Learning Strategies}
When the dictionary $D$ is unknown, the optimization space expands into a joint matrix factorization problem across a data ensemble $Y = [\mathbf{y}_1, \mathbf{y}_2, \dots, \mathbf{y}_N]$. The~K-SVD algorithm \cite{ela06} is a popular  iterative dictionary learning framework that generalizes K-means clustering by alternating between an OMP-based sparse coding phase to recover latent coefficients under cardinality constraints, and a dictionary update phase that employs Singular Value Decomposition (SVD) to sequentially optimize individual dictionary atoms and their non-zero coefficients simultaneously.
Other approaches to dictionary learning include Method of Optimal Directions (MOD), Online Dictionary Learning (ODL), Non-Negative Matrix Factorization (NMF), and Task-Driven Dictionary Learning.

\subsubsection*{Low-Rank Models}
Sparse modeling maps a vector onto an overcomplete matrix. Low-rank matrix approximation extends this intuition by minimizing the structural rank of a data matrix $\mathbf{X}$. A low-rank matrix is a matrix that is sparse in the singular value domain. The non-convex $\text{rank}(\mathbf{X})$ constraint is relaxed using its convex envelope, the Nuclear Norm ($\|\mathbf{X}\|_* = \sum \sigma_i(\mathbf{X})$), which minimizes the sum of the singular~values. 
\vspace{-18pt}

\subsection*{Non-Local Priors: Patch-Based Self-Similarity}

Traditional image priors, such as NMNV, TV, and sparse modeling, are local: they assume a pixel is most like its immediate neighbors. Non-local (NL) priors are built on the observation that natural images often contain patches with repeating patterns. If there is a specific texture on a building’s brick or a pattern in a fabric, chances are that exact same pattern exists elsewhere in the image. 

Non-Local Means (NL-Means) proposed by Buades et al. \cite{bua05} was the first practical implementation that demonstrated "non-local averaging" across the image was far superior to "local averaging."  A non-local estimate $\hat{x}(i)$ at pixel $i$ is a weighted average of all pixels $j$ in the image:
$$\hat{x}(i) = \frac{1}{Z(i)} \sum_{j \in \Omega} w_{\text{NLM}}(i, j) y(j)$$
The weights $w_{\text{NLM}}(i, j)$ are determined by the structural similarity between the patches $\mathcal{P}_i$ and $\mathcal{P}_j$ centered at pixels $i$ and $j$, usually computed using a Gaussian kernel of the Euclidean distance:
$$w_{\text{NLM}}(i, j) = \exp\left( -\frac{\|\mathcal{P}_i - \mathcal{P}_j\|_{2, a}^2}{h^2} \right) \vspace{-5pt} $$
where $\|\cdot\|_{2, a}^2$ denotes a Gaussian-weighted Euclidean distance across the patch footprint, $h$ acts as a filtering parameter that controls the "strictness" of the similarity, and $Z(i) = \sum_{j \in \Omega} w_{\text{NLM}}(i, j)$ is a normalizing partition function.

The conceptual precursor to this paradigm is the Bilateral Filter, which smooths images by combining spatial domain closeness with radiometric (intensity) similarity:
$$w_{\text{BF}}(i, j) = \exp\left( -\frac{\|i - j\|_2^2}{2\sigma_s^2} \right) \exp\left( -\frac{\|y_i - y_j\|_2^2}{2\sigma_r^2} \right) \vspace{-5pt} $$
where $i$ and $j$ represent pixel coordinates, $y$ represents pixel intensities, $\sigma_s$ controls spatial decay, and $\sigma_r$ controls radiometric decay.

NL priors are renowned for restoring highly complex, periodic textures. They are equally potent when regularizing edges. A local estimator can only smooth along an immediate neighborhood of an~edge. In contrast, an NL prior identifies structurally similar edge-segments across non-adjacent regions, effectively averaging noise along the entire directional trajectory of the edge without inducing spatial blur.

While a global search  over the entire image domain $\Omega$ is theoretically optimal for infinite ensembles, practical implementations restrict the search space to a localized bounding window $\mathcal{S}_i$ of size $S \times S$ centered around the target pixel $i$. This bounded search strategy delivers two critical advantages:\\
1) Mitigation of Structural Hallucinations: Searching the entire image introduces a high probability of encountering "spurious matches"—unrelated distant patches that exhibit low Euclidean distance due to noise realizations. Restricting the search space to $\mathcal{S}_i$ ensures that matches remain anchored to contextually relevant structures, yielding superior peak signal-to-noise ratio (PSNR) and fewer visual artifacts.\\
2) Computational Complexity Reduction: An exhaustive global search over an image with $N$ pixels requires an intractable computational complexity of $\mathcal{O}(N^2 \cdot |\mathcal{P}|)$, where $|\mathcal{P}|$ is the patch cardinality. By bounding the search to a localized window of size $|\mathcal{S}|$, the complexity scales linearly with the image size to $\mathcal{O}(N \cdot |\mathcal{S}| \cdot |\mathcal{P}|)$, making the framework viable for high-resolution processing pipelines.

{\it Bayesian Formulation of Non-Local Regularization:}
We can formalize the non-local prior within a MAP-based Bayesian framework. Let $x$ represent the~latent clean image. We treat the non-local weights $w(i,j)$ as structural hyperparameters that define a non-local graph manifold over the image.
The non-local energy functional can be cast as a Graph Dirichlet Form, which functions as the log-prior distribution over the image space:
$$p(\mathbf{x}) \propto \exp\left( -\lambda R_{\text{NL}}(\mathbf{x}) \right) = \exp\left( -\frac{\lambda}{2} \sum_{i \in \Omega} \sum_{j \in \mathcal{S}_i} w(i, j) (x_i - x_j)^2 \right)$$
By applying the negative logarithm, this prior seamlessly pairs with a standard Gaussian noise likelihood model ($\mathbf{y} = \mathbf{x} + \mathbf{n}$), yielding the  variational optimization problem:
$$\hat{\mathbf{x}}_{\text{MAP}} = \arg\min_\mathbf{x} \left\{ \frac{1}{2\sigma^2} \|\mathbf{x} - \mathbf{y}\|_2^2 + \frac{\lambda}{2} \sum_{i \in \Omega} \sum_{j \in \mathcal{S}_i} w(i, j) (x_i - x_j)^2 \right\}$$
\vspace{-4pt}

{\it The Non-Local Laplacian Operator}:
Taking the derivative of the non-local log-prior with respect to $x_i$ yields the functional's score, which corresponds to the Non-Local Laplacian operator ($\Delta_{\text{NL}}$):
$$\left( \Delta_{\text{NL}} \mathbf{x} \right)_i = \sum_{j \in \mathcal{S}_i} w(i, j) (x_j - x_i)$$
The corresponding Euler-Lagrange equation for the MAP inference steps satisfies:
$$\frac{1}{\sigma^2}(x_i - y_i) - \lambda \left( \Delta_{\text{NL}} \mathbf{x} \right)_i = 0$$
This Bayesian formulation reveals that the non-local prior acts as an adaptive, data-driven diffusion process. In homogeneous regions or along consistent geometric edges, the weights $w(i,j)$ remain large, driving strong non-local smoothing. Conversely, at isolated structures where no matching patches exist within $\mathcal{S}_i$, the weights collapse to zero, completely deactivating the regularizer to preserve the rare detail.

{\it A Taxonomy of Non-Local Frameworks}:
Algorithms based on NL image priors were the top performers before the deep learning era. The field of non-local image processing has expanded beyond basic patch-averaging filters into sophisticated sparse and low-rank structural group models. Examples include:

1) Block-Matching and 3D Filtering (BM3D) \cite{dab07}, which combines patch self-similarity with transform-domain sparsity, was the "gold standard" of image denoising for over a decade. It consists of three stages: (a) Grouping: Find similar patches (Block-Matching) and stack them into a 3D array (a "3D block").\\ (b)~Collaborative Filtering: Apply a 3D transform (like DCT or Wavelets) to the~stack.  Because the~patches are similar, the signal is highly sparse in 3D, while the noise is spread out. 
(c) Shrinkage and Inversion: Apply 3D hard-thresholding and empirical Wiener filtering
and transform back to the 2D image.

2) Non-Local Total Variation (NL-TV) replaces the local spatial gradient vector with a non-local gradient operator defined as $\nabla_{\text{NL}} x(i, j) = (x_j - x_i)\sqrt{w(i,j)}$. Minimizing the $\ell_1$-norm of this non-local gradient matrix allows the optimizer to enforce piecewise flatness across non-contiguous but structurally similar regions, eliminating the localized staircasing artifacts typical of standard ROF models.

3) Weighted Nuclear Norm Minimization (WNNM) \cite{gu14} treats the stack of similar patches as a low-rank matrix and minimizes its nuclear norm (sum of singular values) to enforce structural consistency. WNNM assigns inversely proportional weights to different singular values. Larger singular values (which represent dominant structural components) are penalized less, while smaller singular values (which represent noise and residual artifacts) are penalized more, yielding state-of-the-art preservation of subtle textures.

\hypertarget{denoiser-prior}{%
\section*{Denoiser as a Prior: Bridge to Learning Image Prior from Big Data}
\label{denoiser-prior}}

Rather than hand-crafting an analytical prior (such as TV or sparsity), denoiser as prior paradigm employs denoisers—ranging from classical patch-based collaborative filters (BM3D) to deep convolutional neural networks (DnCNN)—as implicit priors.
Denoiser as prior paradigm employs a "divide and conquer" strategy leveraging the split form of the cost function:
$$E(\mathbf{x}) = D(\mathbf{x},\mathbf{y}) + R(\mathbf{x})$$
where the "data fidelity" term $D(\mathbf{x},\mathbf{y})$ is smooth/differentiable (e.g., least squares) and a closed-form or gradient-descent solution exists, and the "regularizer" term $R(\mathbf{x})$ typically has "sharp corners" (e.g., when we employ $L_1$ or TV regularization) and cannot be differentiated. This separation of data-fidelity term and the prior aligns well with Proximal Algorithms and Variable Splitting Algorithms that are commonly used in the optimization literature to handle non-differentiable priors.
\vspace{-6pt}

\subsection*{Proximal Algorithms} 
The~word "proximal" comes from "proximity." These algorithms replace "gradient step" (in the direction of the~negative gradient) with a generalized projection operator ("proximal operator") jumping to a point nearby an input vector $\mathbf{v}$ that minimizes:
$$\text{prox}_{\lambda R}(\mathbf{v}) = \arg\min_\mathbf{x} \left( R(\mathbf{x}) + \frac{1}{2\lambda} \|\mathbf{x} - \mathbf{v}\|^2 \right)$$ 
Parikh and Boyd \cite{par14} is a highly-cited reference for Proximal Algorithms. 
Methods foundational to proximal optimization include Proximal Gradient Descent (PGD), and proximal splitting techniques such as the Alternating Direction Method of Multipliers (ADMM) and Half-Quadratic Splitting (HQS).

The Proximal Gradient Descent algorithm alternates between two simple steps:

Step 1: The Gradient Step (Handling the smooth part)
First, we ignore the non-smooth part $R(\mathbf{x})$ and take a standard gradient descent step using only $D(\mathbf{x}, \mathbf{y})$, which moves the solution closer to the data:
$$\mathbf{v}_k = \mathbf{x}_k - \alpha \nabla D(\mathbf{x}_k,\mathbf{y})$$

Step 2: The Proximal Step (Handling the non-smooth part)
Now, we apply the "Proximal Operator" of $R(\mathbf{x})$ to project the result $\mathbf{v}_k$ onto the desired structure (e.g., clipping small values to zero for sparsity)
$$\mathbf{x}_{k+1} = \text{prox}_{\alpha R}(\mathbf{v}_k)$$

\noindent 
A popular example of a proximal operator is Soft Thresholding. If $R(\mathbf{x}) = \Vert{}\mathbf{x}\Vert{}_1$ ($L_1$ norm), which is used for sparse coding or Lasso, the proximal operator $\text{prox}_{\lambda R}(\mathbf{v})$ applies soft thresholding component-wise:$$\left[ \text{prox}_{\lambda R}(\mathbf{v}) \right]_i = \text{sign}(v_i) \max(\vert{}v_i\vert{} - \lambda, 0), \quad i = 1, \dots, n$$
If the magnitude of an entry $\vert{}v_i\vert{}$ is smaller than $\lambda$ (noise), it is clipped to 0; if it is larger (signal), it is shrunk toward 0 by $\lambda$.

Iterative Shrinkage-Thresholding (ISTA) is a specific instance of PGD. Because the proximal operator of the $L_1$ norm is the soft-thresholding (or shrinkage) operator, PGD applied to the Lasso problem, where $R(x)$ is an $L_1$-regularizer, is called ISTA. The connection between "soft-thresholding" and Lasso problems was formalized by Daubechies {\it et al.} \cite{dau04}. 
Fast ISTA~(FISTA)~\cite{bec09} is an accelerated version of ISTA that converges at a rate of $O(1/k^2)$ using a Nesterov-like momentum term, while ISTA converges at a rate of $O(1/k)$, which means the difference between the objective value at iteration $k$ and the true minimum decreases at a rate proportional to $1/k^2$ and $1/k$, respectively. FISTA has been used for sparse coding and image restoration for over a decade.  number of iterations.
\vspace{-6pt}

\subsection*{Variable Splitting Algorithms}
HQS and ADMM are advanced "Variable Splitting" algorithms. They are used to solve optimization problems when the argument of the~"spiky" function involves a linear transformation (like a blur kernel or a matrix), where simple proximal gradient descent cannot be used.

Recall that ISTA solves $D(\mathbf{x},\mathbf{y}) + R(\mathbf{x})$. But real-world imaging problems often look like: \vspace{-5pt}
$$\min_{\mathbf{x}} D(\mathbf{x},\mathbf{y}) + R(\mathbf{Lx})$$
where $\mathbf{L}$ is a linear operator, which maps the image $\mathbf{x}$ into a domain where it becomes sparse (e.g.,~derivative operator for TV).
Because $\mathbf{L}$ is inside the function $R$, we~can't just threshold~$\mathbf{x}$ to compute the~proximal operator, because the sparsity constraint applies to $\mathbf{Lx}$, not $\mathbf{x}$ itself.

Both HQS and ADMM invent a new variable $\mathbf{z}$ to replace $\mathbf{Lx}$. We rewrite the problem as: \vspace{-3pt}
$$\min_{\mathbf{x}, \mathbf{z}} D(\mathbf{x},\mathbf{y}) + R(\mathbf{z}) \quad \text{subject to} \quad \mathbf{Lx} = \mathbf{z}\vspace{-5pt}$$
where two variables $\mathbf{x}$ and $\mathbf{z}$ are linked by a constraint. The solution consists of two steps: \\
(i) $\mathbf{x}$-step: Solves an inversion problem with quadratic loss (usually has closed-form solution). \\
(ii) $\mathbf{z}$-step: Solves the non-smooth problem using a Proximal Operator (easy because $\mathbf{L}$ is gone).

HQS was introduced to handle non-convex regularizers in image restoration. It relaxes the strict equality constraint $\mathbf{Lx} = \mathbf{z}$ by adding a penalty if they are different. The additive form of HQS minimizes~\cite{gem95} \vspace{-5pt}
$$\text{Cost} = D(\mathbf{x},\mathbf{y}) + R(\mathbf{z}) + \frac{\mu}{2} \| \mathbf{Lx} - \mathbf{z} \|^2 \vspace{-5pt} $$ \vspace{-5pt}
The HQS Algorithm: \\
- Update $\mathbf{x}$: Given $\mathbf{z}_k$, solve a quadratic inverse problem (hence, it is called "half-quadratic"): \vspace{-3pt}
$$\mathbf{x}_{k+1} = \arg \min_{\mathbf{x}} \big( D(\mathbf{x},\mathbf{y}) + \frac{\mu}{2} \| \mathbf{Lx} - \mathbf{z}_k \|^2 \big )$$
- Update $\mathbf{z}$: Apply the Proximal Operator of $R$. Minimizing $R(\mathbf{z}) + \frac{\mu}{2}\Vert{}\mathbf{Lx}_{k+1} - \mathbf{z}\Vert{}^2$ yields: \vspace{-5pt}
$$\mathbf{z}_{k+1} = \text{prox}_{\frac{1}{\mu}R}(\mathbf{Lx}_{k+1})\vspace{-5pt} $$ 
- Increase $\mu$: Make the penalty stricter in the next iteration.

HQS is simple and fast. However, to get an exact solution, $\mu$ must go to infinity, which can cause numerical instability.
ADMM uses an augmented Lagrangian. It keeps the~penalty term from HQS, but adds a "Lagrange multiplier" ($\mathbf{u}$) that tracks the~error between $\mathbf{Lx}$ and $\mathbf{z}$. \\
The ADMM Algorithm (3 Steps): \\
- Update $\mathbf{x}$ (Inversion): Minimize $D(\mathbf{x},\mathbf{y})$ while trying to match $\mathbf{z} - \mathbf{u}$. \vspace{-3pt} 
$$\mathbf{x}_{k+1} = \arg\min_\mathbf{x} \left( D(\mathbf{x},\mathbf{y}) + \frac{\mu}{2} \Vert{} \mathbf{Lx} - \mathbf{z}_k + \mathbf{u}_k \Vert{}^2 \right)$$
- Update $\mathbf{z}$ (Denoising/Proximal Step): Apply Proximal Operator to $\mathbf{Lx} + \mathbf{u}$.
\vspace{-7pt}
$$\mathbf{z}_{k+1} = \text{prox}_{\frac{1}{\mu}R}(\mathbf{Lx}_{k+1} + \mathbf{u}_k) \vspace{-2pt}$$
- Update $\mathbf{u}$ (Dual Correction): Update the multiplier based on the error.\vspace{-7pt}
$$\mathbf{u}_{k+1} = \mathbf{u}_k + (\mathbf{Lx}_{k+1} - \mathbf{z}_{k+1}) \vspace{-7pt}$$
Unlike HQS, ADMM~guarantees convergence to the exact solution without needing the~penalty parameter to go to infinity. 
Boyd et al. \cite{boy11} is a highly-cited guide to ADMM. 
\vspace{-8pt}

\subsection*{Plug-and-Play (PnP) Framework} 
\label{pnp}

The Plug-and-Play (PnP) framework allows solving complex inverse problems by using an off-the-shelf denoiser as a "prior," even if that denoiser was never explicitly designed for the task at hand. 

In the proximal step, we normally solve \vspace{-2pt}: 
$$\text{prox}_{\lambda R}(\mathbf{v}) = \arg\min_\mathbf{x} \frac{1}{2}\|\mathbf{x} - \mathbf{v}\|^2 + \lambda R(\mathbf{x})$$
This specific optimization is what a denoiser does. 
Venkatakrishnan {\it et al.} \cite{ven13} coined the term "Plug-and-Play Priors," demonstrating how in the "z-step" of ADMM, the sophisticated proximal operator of a handcrafted prior (like TV) can be replaced with a simple, state-of-the-art denoiser,
$\mathbf{x}_{k+1} = \text{Denoiser}(\mathbf{v}_k)$.
They used the BM3D (Block-Matching and 3D Filtering) denoiser. 
Following their work, ADMM and HQS have become practical methods to solve inverse problems using PnP priors. 

While \cite{ven13} demonstrated that the PnP framework worked empirically, they did not provide a 
proof of convergence for the algorithm.
Chan {\it et al.} \cite{cha16} provided the first theoretical grounding for~PnP. They established conditions (such as bounded denoiser error) under which the~PnP-ADMM algorithm is guaranteed to converge to a fixed point and implicitly contains the prior of the image distribution.
\vspace{-10pt}

\subsection*{The Deep Learning Bridge (PnP-CNN)}
As deep learning began to dominate denoising, researchers realized they could "plug in" a learned denoiser into the PnP framework to get the best of both worlds: the physics-based accuracy of traditional optimization and the speed and power of deep learning.
Zhang et al. has \cite{zha17} demonstrated that a set of CNN denoisers (DnCNN), trained at different noise levels, could be "plugged" into  HQS or ADMM
outperforming traditional PnP methods while being significantly faster.
\vspace{-6pt}

\subsection*{Regularization by Denoising (RED)} 
\label{red}
While PnP originated as an engineering heuristic, RED~\cite{rom17}, introduced by Romano, Elad, and Milanfar, aimed at presenting a mathematical framework to justify using a denoiser inside the optimization loop. The  original RED formulation claims to construct an explicit regularization functional $R(\mathbf{x})$ directly from a user-defined, black-box denoising function $f(\mathbf{x})$. The proposed regularizer takes the form:
$$R(\mathbf{x}) = \frac{1}{2} \mathbf{x}^T \Big( \mathbf{x} - f(\mathbf{x}) \Big)$$
This functional penalizes the difference between the latent image $\mathbf{x}$ and its denoised counterpart $f(\mathbf{x})$, acting as a structural mismatch penalty. When paired with a data-fidelity term under the MAP inference objective, the optimization problem is stated as:
$$\min_\mathbf{x} \Big\{ E(\mathbf{x}) = D(\mathbf{x}, \mathbf{y}) + \frac{\lambda}{2} \mathbf{x}^T \big( \mathbf{x} - f(\mathbf{x}) \big) \Big\}$$
where $D(\mathbf{x}, \mathbf{y})$ is the log-likelihood data fidelity term (e.g., $D(\mathbf{x}, \mathbf{y})=\frac{1}{2}\|\mathbf{y} - \mathbf{Ax}\|_2^2$ for Gaussian noise). To minimize this energy functional via gradient descent or proximal splitting methods, Romano et al. asserted that the~gradient of the regularizer simplifies to the standard denoising residual:
$$\nabla R(\mathbf{x}) = \mathbf{x} - f(\mathbf{x})$$
This expression allowed RED to bypass the complex derivatives of the underlying denoiser, leading to highly efficient algorithms such as RED-Gradient Descent, RED-Fixed Point, and RED-ADMM.

Despite the empirical success of RED, Reehorst and Schniter \cite{ree19} subsequently demonstrated that the~interpretation of RED as minimizing an explicit regularization function is, in general, mathematically untenable. They showed that the exact gradient of the proposed regularizer $R(\mathbf{x}) = \frac{1}{2}\mathbf{x}^T \mathbf{x} - \frac{1}{2}\mathbf{x}^T f(\mathbf{x})$ by applying standard vector calculus is actually:
$$\nabla R(\mathbf{x}) = \mathbf{x} - \frac{1}{2} f(\mathbf{x}) - \frac{1}{2} \big[ J_f(\mathbf{x}) \big]^T \mathbf{x}$$
where $J_f(\mathbf{x}) \in \mathbb{R}^{n \times n}$ represents the Jacobian matrix of the denoiser $f(\mathbf{x})$, defined element-wise as $[J_f(\mathbf{x})]_{i,j} = \frac{\partial f_i(\mathbf{x})}{\partial x_j}$. For the original RED gradient assumption $\nabla R(\mathbf{x}) = \mathbf{x} - f(\mathbf{x})$ to hold, the denoiser must satisfy two restrictive mathematical conditions simultaneously:
1) The Jacobian matrix must be perfectly symmetric everywhere, meaning $J_f(\mathbf{x}) = [J_f(\mathbf{x})]^T$.
2) The denoiser must satisfy the condition $J_f(\mathbf{x})\mathbf{x} = f(\mathbf{x})$, meaning $f(\mathbf{x})$ behaves as a locally homogeneous function of degree one.

In optimization theory, these conditions restrict $f(\mathbf{x})$ to behaving identically to a proximal operator.
Reehorst and Schniter \cite{ree19} showed that practical denoisers do not possess a symmetric Jacobian. The~patch-matching weights in algorithms such as NL-Means and BM3D depend non-linearly on the~input image $\mathbf{x}$. Taking the derivative of the patch weights with respect to the input intensities breaks Jacobian symmetry completely. Likewise, convolutional neural networks, such as DnCNN, UNet, utilize non-linear activation functions (e.g., ReLU, GeLU) and data-dependent normalizations. The composition of these layers yields a highly asymmetric Jacobian matrix $J_f(\mathbf{x}) \neq [J_f(\mathbf{x})]^T$.

Because practical denoisers violate Jacobian symmetry, the pseudo-gradient step $\mathbf{x} - f(\mathbf{x})$ in the~RED literature is not the gradient of an actual scalar objective function. Consequently, running a RED algorithm does not trace a path down an energy landscape; instead, it executes an iterative path through a non-conservative vector field, meaning proofs of variational optimization convergence do not apply. Yet~the~RED algorithms remain highly powerful. To resolve this paradox, the literature has abandoned the explicit MAP framework in favor of alternative sound mathematical interpretations of RED.

\section*{Tweedie’s Formula and MMSE Denoiser as Score Estimator}

While the integration of off-the-shelf denoisers into iterative solvers, such as the PnP and RED frameworks, was initially motivated by the intuitive concept of the "denoiser as a prior," it possesses a rigorous mathematical foundation that links classical signal processing to modern generative modeling. This bridge is established through Tweedie’s formula and the principles of denoising score matching. 

Let us consider a noisy observation $\mathbf{y} = \mathbf{x} + \mathbf{n}$, where $\mathbf{x}$ is the clean image drawn from a prior distribution $p(\mathbf{x})$, and $\mathbf{n} \sim \mathcal{N}(0, \sigma^2 \mathbf{I})$ is additive white Gaussian noise. Tweedie’s formula (originally rooted in empirical Bayes \cite{efr11}) demonstrates that the Minimum Mean Square Error (MMSE) estimator for the clean image, given by the conditional expectation $E[\mathbf{x}|\mathbf{y}]$, can be expressed in terms of the noisy observation and the gradient of its log-probability density function: \vspace{-4pt}
$$E[\mathbf{x}|\mathbf{y}] = \mathbf{y} + \sigma^2 \nabla_{\mathbf{y}} \log p(\mathbf{y}) \vspace{-4pt}$$
where the term $\nabla_{\mathbf{y}} \log p(\mathbf{y})$ is known as the score function of the smoothed data distribution. 
By~rearranging this formula, we uncover a profound identity:
$$\nabla_{\mathbf{y}} \log p(\mathbf{y}) = \frac{E[\mathbf{x}|\mathbf{y}] - \mathbf{y}}{\sigma^2}$$
This identity proves that any optimal MMSE denoiser $f(\mathbf{y}) \approx E[\mathbf{x}|\mathbf{y}]$ inherently acts as a score estimator~\cite{rap11}. This expression holds strictly under the assumption of an optimal MMSE denoiser $f(\mathbf{y})$ operating on a Gaussian-smoothed data distribution $p_\sigma(\mathbf{y})$. Note that if a denoiser is truly the optimal MMSE denoiser, a symmetric Jacobian is guaranteed. Consequently, solving an inverse problem no longer requires an explicit algebraic formulation of a prior; it merely requires a forward pass through an optimal MMSE denoiser to compute the update trajectory.
When we train a deep convolutional neural network~(CNN) to remove Gaussian noise using the mean squared error loss, the network is implicitly learning to approximate the~score function of the data manifold. 
\vspace{-6pt}

\subsection*{Interpretation of PnP and RED as Score-Matching by Denoising (SMD)}


Reehorst and Schniter \cite{ree19} proposed replacing explicit regularization with Score-Matching by Denoising (SMD). Drawing from Tweedie’s formula and kernel density estimation, SMD proves that the~expression $\mathbf{x} - f(\mathbf{x})$ represents the score of the data distribution under a MMSE denoiser. Thus, RED algorithms do not minimize an explicit cost function; rather, they perform a root-finding task that matches the score of the underlying data manifold.
\vspace{-10pt}

\subsection*{Transition from Classical to Generative Image Models - Explicit vs. Implicit Priors}

The overarching theme of the evolution of image modeling is the transition from early handcrafted image priors defined analytically via regularizers $\mathcal{R}(\mathbf{x})$ used in  point estimation to implicit (data-driven) priors defined by  $\mathcal{R}(\mathbf{x}) = -\log p_\sigma(\mathbf{x})$, whose gradients are computed operationally by an algorithmic denoiser $f(\mathbf{x})$, where $p_\sigma(\mathbf{x})$ denotes the noise-perturbed distribution, leveraging Tweedie’s formula.

We must emphasize that finding a point MAP estimate via optimization vs. Posterior Sampling via generative/foundation image models are fundamentally different solutions. The "unification" does not mean they will lead to the same solution. Rather, they are unified because they share the same mathematical engine to represent the image prior. Once we have a minimum mean square error (MMSE) denoiser to estimate the Stein score $\nabla_\mathbf{x} \log p_\sigma(\mathbf{x})$, the unification lies in the fact that we can use this same neural network to either optimize (MAP) or simulate (sample).
\vspace{-12pt}

\subsection*{Common Basis of Generative Image Models: Learning High-dimensional Probability Densities}
All generative image models (posterior sampling) address learning high-dimensional probability densities, $p(\mathbf{x})$, to generate new data. However, they have different architectures and different training objectives. VAEs and GANs perform sampling in one step, while DDPM, score- and flow-based models sample along a trajectory.
The  VAE consists of a learned encoder and a learned generative decoder, synthesized via a low-variance reparameterized bottleneck to maximize a normalized density lower bound~(ELBO).
The Diffusion Model 
is a multi-step, hierarchical latent variable model that uses score matching for optimization. 
Its architecture is defined by a pre-defined forward process and a single, learned recursive component for the reverse (sampling) process. 
Generative image models include:

1. DDPM and Score-Based SDEs: While early diffusion models (DDPM by Ho et al.) were described by variational inference and Markov chains, Y. Song et al. unified them via Score-Based Generative Modeling through SDEs. A single network is trained across a continuous spectrum of noise levels $t \in [0, T]$ to estimate the time-dependent score field $\nabla_{\mathbf{x}_t} \log p_t(\mathbf{x}_t)$. By adding controlled noise and stepping along the score vector field using Langevin Dynamics, we can start from pure white noise at $t=T$ and walk systematically down into a highly realistic, high-probability image at $t=0$.

\begin{figure}[t!]
\begin{center}
\includegraphics[width=0.96\textwidth]{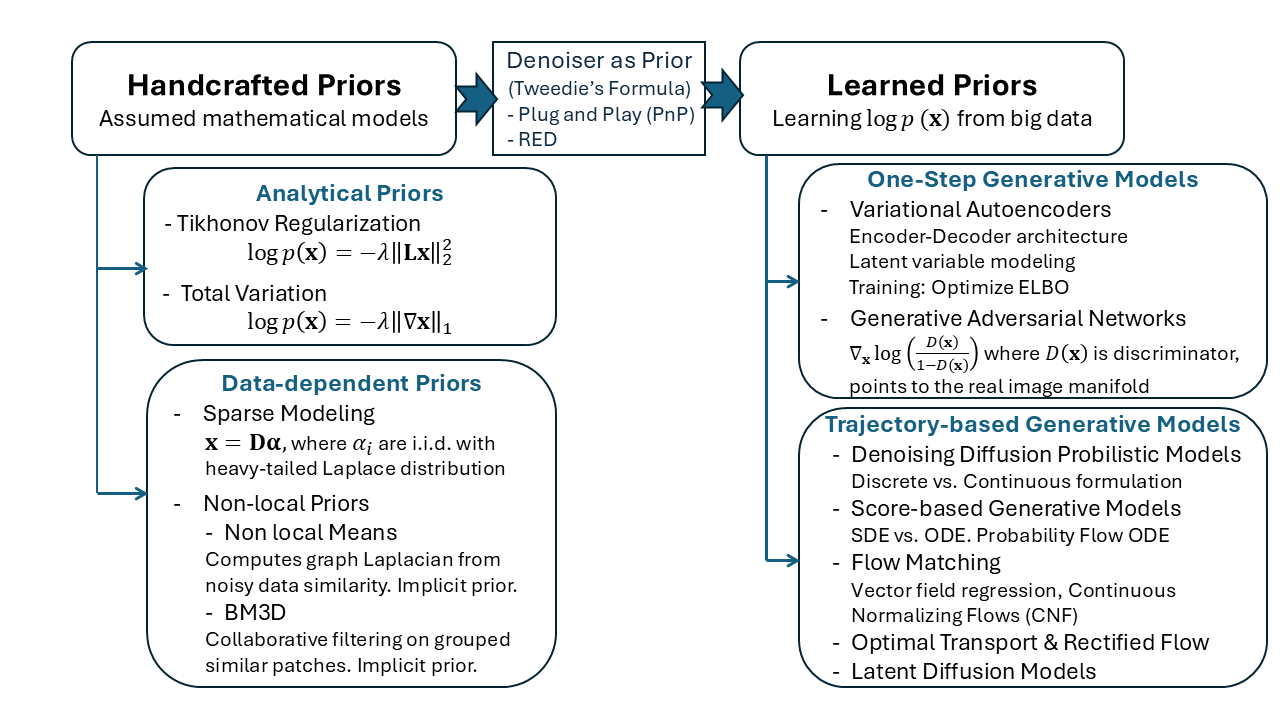} \vspace{-28pt}
\caption{Unification of image modeling from handcrafted priors to learned foundation image models based on Bayesian framework.} \vspace{-25pt}
\end{center}
\label{fig:sonn-sr}
\end{figure}

2. Flow Matching and Rectified Flows: While diffusion models navigate the score field using curved, stochastic paths, Flow Matching and Rectified Flows parameterize the problem as a deterministic velocity vector field. Instead of modeling the score directly, they learn a vector field that generates straight, linear trajectories between noise and data, resulting in significantly faster and more stable sampling.

3. Variational Autoencoders (VAE): VAEs avoid modeling the complex image-space score field directly by mapping the data to a low-dimensional, latent space $\mathbf{z}$ governed by an explicit prior like a standard Gaussian $p(\mathbf{z})$. However, both the generative model (the decoder, $p(\mathbf{x}|\mathbf{z})$) and the optimal inference process (the posterior, $p(\mathbf{z}|\mathbf{x})$) are intractable unknowns. VAEs minimize ELBO with the reparameterization trick to find low-variance gradients because it is the only computationally stable way to learn a decoder $p(\mathbf{x}|\mathbf{z})$, where the optimal $\mathbf{z}$ itself is a moving target that must also be learned.
However, the gradient of the VAE’s reconstruction loss with respect to the input image acts as an empirical approximation of the~score field, pushing corrupted inputs back toward the manifold determined by the latent bottleneck.

4. Latent Diffusion Models (LDM): LDMs are the ultimate synthesis of Variational Inference and Denoising Score Matching. A VAE is trained first to learn a normalized $p(\mathbf{x}|\mathbf{z})$ with tractable ELBO. Then, a diffusion model is trained within the latent space $\mathbf{z}$ using score matching to learn the~prior $p(\mathbf{z})$.

5. Generative Adversarial Networks (GANs): GANs bypass probability density estimation, framing generation as a two-player game. The gradient of the log-odds ratio: $\nabla_\mathbf{x} \log \left( \frac{D(\mathbf{x})}{1-D(\mathbf{x})} \right) = \nabla_\mathbf{x} \log p_{\text{real}}(\mathbf{x}) - \nabla_\mathbf{x} \log p_{\text{fake}}(\mathbf{x})$, where $D(\mathbf{x})$ is the discriminator, pushes generated images toward the high-density "real" image manifold.

The various stages of the evolution of image modeling from handcrafted image priors in pre 1980's to priors learned from big data is summarized in Figure 2. Clearly, the emergence of today's diffusion-based foundation models is not a disjointed leap, but rather a continuous evolution centered around learning and traversing the score of the image manifold.
\vspace{6pt}

\vspace{-3pt}

\section*{Generative Image Models: Variational Autoencoder}
\label{vae-prior}
Our goal is to model the distribution $p(\mathbf{x})$ of our data $\mathbf{x}$. Variational autoencoders~(VAE) assume $\mathbf{x}$ is generated by some lower-dimensional latent (unobserved) variables $\mathbf{z}$, that is, $\mathbf{x}$~is a nonlinear function of some latent variables $\mathbf{z}$. 
We~wish to answer: "Given an image $\mathbf{x}$, what is the probability that the latent variable is $\mathbf{z}$?" That~is, we want to find the posterior $p(\mathbf{z}|\mathbf{x})$. 
Using Bayes' Theorem, we have
$$p(\mathbf{z}|\mathbf{x}) = \frac{p(\mathbf{x}|\mathbf{z})p(\mathbf{z})}{p(\mathbf{x})}$$ \vspace{-6pt}
where $p(\mathbf{x})$, called the Evidence, is given by
$$p(\mathbf{x}) = \int p(\mathbf{x}|\mathbf{z})p(\mathbf{z}) \, d\mathbf{z}$$
For most real-world data, such as images, this integral is intractable because it requires searching through all possible (continuous) values of $\mathbf{z}$. Because we can't calculate $p(\mathbf{x})$, we can't calculate $p(\mathbf{z}|\mathbf{x})$.
\vspace{-8pt}

\subsection*{Variational Inference and the Evidence Lower Bound}
VAEs were almost simultaneously introduced by Kingma {\it et al.} \cite{kin14} and Rezende {\it et al.} \cite{rez14}. Variational inference picks a simple distribution~$q(\mathbf{z})$ (such as a Gaussian), and poses minimization of the~difference between our guess $q(\mathbf{z})$ and the true posterior $p(\mathbf{z}|\mathbf{x})$ as a variational problem. The measure of difference between distributions is the~Kullback-Leibler (KL) Divergence $D_{KL}(q(\mathbf{z}) || p(\mathbf{z}|\mathbf{x}))$.

Let's start with the definition of KL Divergence. It is the expectation (average) of the log difference:
\begin{eqnarray}
D_{KL} &=& E_q [\log q(\mathbf{z}) - \log p(\mathbf{z}|\mathbf{x})] 
      = E_q [\log q(\mathbf{z})] - E_q [\log p(\mathbf{z}|\mathbf{x})]
\end{eqnarray}
Since $p(\mathbf{z}|\mathbf{x}) = \frac{p(\mathbf{x}, \mathbf{z})}{p(\mathbf{x})}$, we have
$\log p(\mathbf{z}|\mathbf{x}) = \log p(\mathbf{x}, \mathbf{z}) - \log p(\mathbf{x})$.
Substituting this back into above:
$$D_{KL} = E_q [\log q(\mathbf{z})] - E_q [\log p(\mathbf{x}, \mathbf{z}) - \log p(\mathbf{x})]$$
Notice that $\log p(\mathbf{x})$ does not depend on $\mathbf{z}$.
Therefore, it can be pulled out of the expectation:
Since we want to solve for $\log p(\mathbf{x})$, let's move everything else to the other side:
$$\underbrace{\log p(\mathbf{x})}_{\text{Evidence}} = \underbrace{E_q [\log p(\mathbf{x}, \mathbf{z})] - E_q [\log q(\mathbf{z})]}_{\text{ELBO}} + \underbrace{D_{KL}(q(\mathbf{z}) || p(\mathbf{z}|\mathbf{x}))}_{\text{KL divergence (Error)}}$$
Since the KL divergence is always non-negative,
dropping the KL divergence term yields an inequality:
$$\log p(\mathbf{x}) \ge \text{ELBO}$$
i.e., the ELBO is strictly lower than or equal to the Evidence.
Since the Evidence $\log p(\mathbf{x})$ is fixed (the data is given),  maximizing the ELBO means minimizing the KL Divergence (the error).
From basic probability, $p(\mathbf{x}, \mathbf{z}) = p(\mathbf{x}|\mathbf{z}) \cdot p(\mathbf{z})$. Hence,
$\log p(\mathbf{x}, \mathbf{z}) = \log p(\mathbf{x}|\mathbf{z}) + \log p(\mathbf{z})$.
Substituting this into the definition of ELBO, we have
$$\text{ELBO} = E_q [ (\underbrace{\log p(\mathbf{x}|\mathbf{z}) + \log p(\mathbf{z})}_{\log p(\mathbf{x},\mathbf{z})}) - \log q(\mathbf{z}) ] = \underbrace{E_q [\log p(\mathbf{x}|\mathbf{z})]}_{\text{Likelihood}} + \underbrace{E_q [\log p(\mathbf{z}) - \log q(\mathbf{z})]}_{\text{Latent}}$$
where we regroup the terms to separate the "Likelihood" part from the "Latent" part. In order to express the latent part as KL Divergence, recall: $D_{KL}(q || p) = E_q [ \log q(\mathbf{z}) - \log p(\mathbf{z}) ]$
and observe that the latent part is the negative of the KL Divergence:
\begin{eqnarray}
\text{ELBO} = E_q [\log p(\mathbf{x}|\mathbf{z})] - \underbrace{E_q [\log q(\mathbf{z}) - \log p(\mathbf{z})]}_{D_{KL}} = \underbrace{E_q [\log p(\mathbf{x}|\mathbf{z})]}_{\text{Reconstruction Loss}} - \underbrace{D_{KL}(q(\mathbf{z}) || p(\mathbf{z}))}_{\text{Regularization Loss}}
\end{eqnarray}
The second term is regularization loss, which forces ~the~latent distribution to be close to a prior (usually~$\mathcal{N}(0, \textbf{I})$), preventing the model from ``cheating'' by assigning a unique point to every input. Hence, the latent space of VAEs is regularized and continuous allowing learning the underlying probability distribution of a dataset for new sample generation.
\vspace{-20pt}

\subsection*{The Reparameterization Trick}
In a VAE, the encoder doesn’t output a latent vector; it outputs parameters of a distribution: a mean~$\mu_i$ and a variance $\sigma_i^2$ for each latent dimension. We then sample a latent vector $\mathbf{z}$ from this distribution:
$\mathbf{z} \sim \mathcal{N}\left(\mathbf{\mu}, \text{diag}(\mathbf{\sigma}^2)\right)$
where $\mathbf{\mu} = [\mu_1, \mu_2, \dots, \mu_d]^T$ is the mean vector and $\mathbf{\sigma}^2 = [\sigma_1^2, \sigma_2^2, \dots, \sigma_d^2]^T$ is the~variance vector.
Sampling is a stochastic process and has no gradient. 
Hence, we cannot use the~gradients to improve $\mathbf{\mu}$ and $\mathbf{\sigma}^2$ estimates. 
The reparameterization trick expresses $\mathbf{z}$ as a deterministic function of another random variable $\mathbf{\epsilon}$ with a standard normal distribution $\mathbf{\epsilon} \sim \mathcal{N}(0, \textbf{I})$ as \cite{kin14}: \vspace{-6pt}
$$\mathbf{z} = \mathbf{\mu} + \mathbf{\sigma} \odot \mathbf{\epsilon} \vspace{-4pt} $$
This way, $\mathbf{z}$ becomes a differentiable function of $\mathbf{\mu}$ and~$\mathbf{\sigma}$, and the gradients can flow from the loss function, through the decoder, through the $\mathbf{z}$ calculation, and back into the encoder’s $\mathbf{\mu}$ and $\mathbf{\sigma}$ layers.
\vspace{-10pt}

\subsection*{Training VAE}
Given a training set of images $\mathbf{X} = \{\mathbf{x}^{(1)}, \mathbf{x}^{(2)}, \dots, \mathbf{x}^{(N)}\}$, the workflow to train an Encoder ($q_\phi$) and a Decoder ($p_\theta$) simultaneously  for a mini-batch of $B$ training images $\{\mathbf{x}^{(b)}\}_{b=1}^B$ is as follows:

\paragraph{Encode}
For each image $\mathbf{x}^{(b)}$ in the batch, the encoder predicts the mean vector $\boldsymbol{\mu}^{(b)} \in \mathbb{R}^J$ and log-variance vector $\log (\boldsymbol{\sigma}^{(b)})^2 \in \mathbb{R}^J$ of the latent distribution, where J is the dimension of the latents.

\paragraph{Reparameterization Trick (Sampling)}
To allow backpropagation through stochastic nodes, sample noise $\boldsymbol{\epsilon}^{(b)} \sim \mathcal{N}(\mathbf{0}, \mathbf{I}_J)$ and construct the latent code via Hadamard (elementwise) product $\odot$: \vspace{-4pt}
$$\mathbf{z}^{(b)} = \boldsymbol{\mu}^{(b)} + \boldsymbol{\sigma}^{(b)} \odot \boldsymbol{\epsilon}^{(b)}$$ \vspace{-28pt}

\paragraph{Decode (Reconstruction)}
Feed each sampled latent vector $\mathbf{z}^{(b)}$ into the Decoder to generate the~reconstructed image:
$\hat{\mathbf{x}}^{(b)} = \text{Decoder}(\mathbf{z}^{(b)})$

\paragraph{Compute the Batch ELBO Objective}
Compute the loss averaged over the mini-batch of size $B$:
$$\mathcal{L}_{\text{batch}} = \frac{1}{B} \sum_{b=1}^B \left( L_{\text{recon}}^{(b)} + L_{\text{KL}}^{(b)} \right) \vspace{-4pt}$$ \vspace{-4pt}
where for sample $b$: \vspace{-4pt}
$$L_{\text{recon}}^{(b)} = \|\mathbf{x}^{(b)} - \hat{\mathbf{x}}^{(b)}\|_2^2 ~~~~\text{and}~~~~ L_{\text{KL}}^{(b)} = -0.5 \sum_{j=1}^J \left( 1 + \log\left((\sigma_j^{(b)})^2\right) - (\mu_j^{(b)})^2 - (\sigma_j^{(b)})^2 \right)$$
\vspace{-14pt}

\paragraph{Backpropagation and Optimization}
Compute the gradients $\nabla_\phi \mathcal{L}_{\text{batch}}$ and $\nabla_\theta \mathcal{L}_{\text{batch}}$ and update weights using an optimizer (e.g., Adam).
\vspace{-12pt}

\subsection*{VAE as a Generative Model}
Once trained, the VAE is used as a generative model to synthesize new samples by following the steps:
\begin{enumerate}
    \item Discard the Encoder. The inference network $q_\phi(\mathbf{z}|\mathbf{x})$ is no longer needed.
    \item Sample a latent vector $\mathbf{z}$ from the standard normal distribution:
    $\mathbf{z} \sim p(\mathbf{z}) = \mathcal{N}(\mathbf{0}, \mathbf{I})$
    \item Pass $\mathbf{z}$ through the generative model $p_\theta(\mathbf{x}|\mathbf{z})$ to yield a generated sample:
    $\hat{\mathbf{x}}_{\text{generated}} = \text{Decoder}(\mathbf{z})$
\end{enumerate}

VAEs provide a principled way to perform non-linear dimensionality reduction while ensuring the~latent space is well-behaved for generation. By balancing reconstruction and regularization losses, they allow for the synthesis of new, plausible data samples.

\hypertarget{gan-gen}{%
\section*{Generative Adversarial Networks }\label{gan-gen}}
Generative Adversarial Networks (GANs) \cite{goo14} are classified as implicit density estimators. Unlike explicit models—such as VAEs, which utilize the~ELBO to approximate data likelihood—GANs do not provide a mechanism to compute the probability of a specific sample. 
The adversarial framework reframes the generative task as a competitive, two-player zero-sum game, where the goal is not to maximize a~likelihood function but to achieve an equilibrium between two competing neural architectures. 
GANs~learn to transform a simple, known distribution (through a latent variable $\mathbf{z}$, typically drawn from a Gaussian or Uniform distribution) into a high-dimensional data manifold. 
\vspace{-6pt}

\subsection*{Architecture of GAN and the Adversarial Metaphor}

The architecture of a GAN consists of two distinct sub-models. The generator $G(\mathbf{z}; \mathbf{\theta}_g)$ maps latent noise $\mathbf{z}$ to a data sample, while the discriminator $D(\mathbf{x}; \mathbf{\theta}_d)$ maps a sample $\mathbf{x}$ to a scalar value representing the probability that $\mathbf{x}$ is real. Both models are implemented as deep neural networks.

The operation of a GAN is as follows: The generator learns to 
generate an image that is indistinguishable from 
a real image and the discriminator learns to detect whether 
a presented image is genuine (from the training set) or a fake (produced by the generator).
Initially, the generator is inept, producing outputs that resemble random noise. However, as the training progresses, the discriminator provides feedback to guide the generator toward the real data distribution. 
This process continues until the generator's outputs are so realistic that the discriminator can no longer distinguish between them, resulting in an equilibrium where the~discriminator's accuracy falls to 50\%, akin to random guessing.
\vspace{-6pt}

\subsection*{Mathematical Foundations: The Minimax Formulation}
The objective function of a GAN is derived from the principles of game theory and information theory.
We first analyze the discriminator's task as a standalone binary classifier. In a binary classification task, for a given sample $\mathbf{x}$ and a true label $y \in \{0, 1\}$, the~binary cross-entropy loss (BCE) is defined as:
\[H(y, \hat{y}) = - [y \log(\hat{y}) + (1 - y) \log(1 - \hat{y})]\]
where $\hat{y}$ is the probability predicted by the discriminator $D(\mathbf{x})$.
For real samples $\mathbf{x}$ from the true distribution $p_{data}$, the~label is $y = 1$. Substituting this into the BCE formula yields:
$L_{real} = -\log(D(\mathbf{x}))$.
For generated samples $G(\mathbf{z})$ from the noise distribution $p_z$, the label is $y = 0$. Substituting this yields:
$L_{fake} = -\log(1 - D(G(\mathbf{z})))$.
To maximize its classification accuracy, the discriminator seeks to minimize the total loss, or equivalently, to maximize the log-probabilities. By taking the expectation over the data and noise distributions, we arrive at the full discriminator objective:
\[\max_D V(D, G) = \mathbb{E}_{\mathbf{x} \sim p_{data}}[\log D(\mathbf{x})] + \mathbb{E}_{\mathbf{z} \sim p_z}[\log(1 - D(G(\mathbf{z})))]\]
\vspace{-12pt}

The generator's goal is the opposite of that of the discriminator. It seeks to minimize the same value function $V(D, G)$, effectively maximizing the probability that the discriminator misclassifies its outputs. 

This leads to the minimax game:
\[\min_G \max_D V(D, G) = \mathbb{E}_{\mathbf{x} \sim p_{data}}[\log D(\mathbf{x})] + \mathbb{E}_{\mathbf{z} \sim p_z}[\log(1 - D(G(\mathbf{z})))]\]
where the discriminator $D$ seeks the maximum value by correctly identifying real and fake data, while the generator $G$ seeks the minimum value by confusing the~discriminator. The term ``minimax'' refers to the search for a saddle point in the parameter space, which corresponds to a Nash Equilibrium in game theory—a state where neither player can improve their outcome by changing their strategy independently.

For any fixed generator $G$, there exists a unique optimal discriminator $D^*_G(\mathbf{x})$ given by
\[D^*_G(\mathbf{x}) = \frac{p_{data}(\mathbf{x})}{p_{data}(\mathbf{x}) + p_G(\mathbf{x})}\]
which represents the posterior probability of the sample being ``real'' given~$\mathbf{x}$, where $p_{data}(\mathbf{x})$ is the~real data density and $p_G(\mathbf{x})$ is the~probability density induced by the generator. That is, the~discriminator is a density ratio estimator that indicates how much more likely a sample is from $p_{data}(\mathbf{x})$ than from $p_G(\mathbf{x})$.
\vspace{-24pt}

\subsection*{Global Optimality and the Jensen-Shannon Divergence}

Substituting the expression for the optimal discriminator $D^*_G$ into the value function $V(G, D^*_G)$ yields the training criterion $C(G)$ that the generator attempts to minimize during the minimax game:
\[C(G) = \mathbb{E}_{\mathbf{x} \sim p_{data}} \left[ \log \frac{p_{data}(\mathbf{x})}{p_{data}(\mathbf{x}) + p_G(\mathbf{x})} \right] + \mathbb{E}_{\mathbf{x} \sim p_G} \left[ \log \frac{p_G(\mathbf{x})}{p_{data}(\mathbf{x}) + p_G(\mathbf{x})} \right]\]
which can be shown to be equivalent to
\[C(G) = 2 D_{JSD}(p_{data} || p_G) - 2 \log(2)\]
where $D_{JSD}(P || Q)$ the Jensen-Shannon Divergence (JSD) between the distributions $P$ and $Q$. Hence, the adversarial process is mathematically equivalent to minimizing the JSD between the~model's induced distribution and the true data distribution.
The JSD is a symmetric and non-negative measure,
which is equal to zero if and only if the two distributions are identical ($p_{data} = p_G$). Therefore, the global minimum of the generator's training criterion is achieved when the generator has perfectly~replicated the~data distribution. At this point, 
$C(G)=-\log(4)$, which corresponds to $D(x) = 0.5$ everywhere.
\vspace{-8pt}

\subsection*{From Vanilla GAN to DCGAN} \vspace{-4pt}
Several architectural refinements have been developed to improve the performance of GANs. The~Deep Convolutional GAN (DCGAN) \cite{rad16} is perhaps the most significant one for high-resolution image synthesis.
It replaced traditional pooling layers with strided convolutions and introduced Batch Normalization into the GAN architecture. Most importantly, it showed that the latent space of a GAN could be used for "vector arithmetic" (e.g., Man with glasses - Man + Woman = Woman with glasses).
\vspace{-8pt}

\subsection*{Practical Training Challenges} \vspace{-4pt}
The simultaneous training of two networks leads to the following practical difficulties:

\subsubsection*{The Vanishing Gradient Problem}
Early in training, the generator's samples $G(\mathbf{z})$ are often poor, and the discriminator can distinguish them from real data with high confidence, which makes $D(G(\mathbf{z}))$ close to 0 and the term $\log(1 - D(G(\mathbf{z})))$ very flat. The gradient of this term with respect to the generator's parameters becomes vanishingly small.
To solve this, \cite{goo14} proposed a non-saturating loss. The generator is trained to maximize $\log(D(G(\mathbf{z})))$ instead of minimizing $\log(1 - D(G(\mathbf{z})))$. While the global equilibrium remains the same, the non-saturating objective provides stronger gradients early in the learning process.

\subsubsection*{Mode Collapse and Lack of Diversity}
Mode collapse is a failure state where the generator learns to produce only a subset of the training data distribution. For instance, a model trained on a dataset~of~diverse faces might collapse into producing only a specific face that happens to fool the~discriminator. Because the discriminator evaluates each sample independently, it does not penalize the generator for lack~of~diversity.
\vspace{-28pt}

\subsection*{Beyond JSD: The Wasserstein GAN (WGAN)} \vspace{-3pt}
WGAN replaced the~traditional Jensen-Shannon Divergence (JSD) cost function with the~Wasserstein distance (Earth Mover's distance), which measures the minimum cost of transporting mass from one distribution to another to solve these issues. Unlike~JSD, the Wasserstein distance is continuous and differentiable even when distributions are disjoint, providing a meaningful gradient that guides the~generator~\cite{arj17}. That is, instead of a "Discriminator" that tells if an image is real or fake~(0~or~1), WGAN uses a "Critic" that provides a continuous score, ensuring the generator always has a gradient to follow. 

While WGAN was theoretically sound, its method of enforcing the "1-Lipschitz" constraint (weight clipping) led to optimization difficulties. 
Gulrajani {\it et al.} \cite{gul17} introduced the Gradient Penalty (GP), which penalizes the model if the gradient norm of the~critic deviates from 1. This became the standard implementation for almost all GAN research that followed, including the famous StyleGAN series.  

\hypertarget{AR-prior}{%
\section*{Autoregressive Image Models}
\label{AR-prior}}

Autoregressive (AR) modeling treats image generation as a sequential task, similar to how Large Language Models predict the next word in a sentence. These models learn to generate pixels (or latent tokens) one by one, typically following a raster-scan order~(top-left to bottom-right). 

The fundamental assumption of an AR model is that the joint probability distribution of all pixels in an  image $\mathbf{x}$ (composed of $n$ pixels) can be decomposed  into a product of conditional probabilities using the~chain rule based on a raster-scan order:
$$p(\mathbf{x}) = \prod_{i=1}^n p(x_i | x_1, \dots, x_{i-1})$$ \vspace{-3pt}
where each pixel $x_i$ is conditioned on all previously generated pixels. Because this modeling is "exact" (it doesn't rely on approximations as in VAEs), AR models are excellent at capturing high-frequency details, though they are computationally expensive due to their sequential nature.
\vspace{-10pt}

\subsection*{Architectures} 

\noindent {\it Phase 1: Causal Convolutions (RNN/CNN)} \\
PixelRNN and PixelCNN \cite{voicml16} use Recurrent Neural Networks (RNNs) and causal masking, respectively, to maintain "causality"—ensuring a pixel only "sees" pixels that came before it.
PixelCNN was presented as a faster alternative to PixelRNN. While PixelRNN was more accurate, PixelCNN replaced the slow, sequential LSTMs with masked convolutions. This allowed the model to be trained in parallel, significantly reducing training time.
Gated PixelCNN \cite{voneur16} replaced simple ReLU activations with gated activation units (similar to LSTMs) and introduced a "stack" of two convolutions (horizontal and vertical) to eliminate the "blind spot."
While PixelCNN \cite{voicml16} proved parallel training was possible for AR models, the~Gated PixelCNN~\cite{voneur16} proved that these models could be "conditioned" (e.g., "Generate a picture of a 5"), which laid the~groundwork for modern text-to-image systems.
The follow-up work PixelCNN++ \cite{sal17} is a refinement that improved efficiency and quality by using a discretized logistic mixture likelihood.
\vspace{2pt}

\noindent {\it Phase 2: Self-Attention (The Transformer)} \\
As Transformers began to dominate NLP, researchers also applied self-attention to images. This allowed models to capture long-range dependencies (e.g., ensuring the left eye matches the right eye) that were difficult for local convolutional windows to manage. Some examples include:
Image Transformer \cite{par18} is the first major application of the Transformer architecture to image synthesis, replacing convolutions with self-attention.
iGPT (Image GPT) \cite{che20} by OpenAI demonstrated that a massive Transformer trained solely to predict the next pixel could learn powerful visual representations without any labels.
\vspace{2pt}

\noindent {\it Phase 3: The Discrete Latent Space}\\
Generating high-resolution images pixel-by-pixel is painfully slow. 
Vector Quantized Variational Autoencoder (VQ-VAE) \cite{vdo17} turns images into a discrete sequence of latent codes, making AR modeling practical for high-resolution image and video generation. 
The VQ-VAE architecture is a standard Encoder-Decoder structure. Encoding takes the image and maps it to a grid of discrete indices (tokens). Once the VQ-VAE is trained, it compresses an image into a grid of integers (e.g., $32 \times 32$ codebook indices). Decoding takes this grid of indices and reconstructs the image in parallel.
This is where the AR model comes in: An AR model (typically a PixelCNN or a Transformer) is trained to predict the latent codes one by one. 
The original DALL-E \cite{ram21} 
utilized a VQ-VAE and a massive AR Transformer to generate images from text, proving that the AR paradigm could scale to multi-modal foundation models.

While diffusion models have largely superseded AR models for pure image synthesis speed, the~AR paradigm remains the~backbone of multimodal intelligence in Foundation Image Models. It allows for a unified "vocabulary" where text, images, and audio are all just tokens in a single, long sequence.

\hypertarget{ddpm-gen}{%
\section*{Trajectory-Based Deep Generative Models: Diffusion, Scores, and Flows}
\label{ddpm-gen}}

Deep generative image modeling has recently undergone a paradigm shift. Moving away from the~discrete, one-shot mappings, such as VAEs and GANs, modern deep generative models (DGM) transform a simple, known prior distribution $p_{\text{prior}}$ (e.g., a standard isotropic Gaussian) into a complex, high-dimensional data distribution $p_{\text{data}}$. These frameworks, including Denoising Diffusion Probabilistic Models (DDPMs), Score-Based Generative Models (SGMs), and Flow Matching~(FM), are collectively designated as trajectory-based generative models, as they explicitly model, optimize, and sample along time-dependent paths of random variables.
\vspace{-6pt}

\subsection*{From Statistical Physics to Signal Processing}
Sohl-Dickstein {\it et al.} \cite{soh15} introduced the concept of using a Markov chain to slowly destroy and then restore structure in a data distribution, inspired by statistical physics. However, the implementation was dense and lacked the visual fidelity required for
applications.
Ho {\it et al.} \cite{ho20} took the physics-inspired foundation and re-engineered it into a DDPM denoising pipeline  
and showed that the model performed significantly better if it predicted the added noise ($\epsilon$) at each step.
\vspace{-6pt}

\subsection*{The Continuous-Time SDEs and Probability Flows}
To unify DDPM, SGM and FM paradigms, we define a continuous time variable $t \in [0, T]$. Without loss of generality, let $t=0$ index the clean data state, $\mathbf{x}_0 \sim p_{\text{data}}(\mathbf{x})$, and let $t=T$ index the fully perturbed noise state, $\mathbf{x}_T \sim p_{\text{prior}}(\mathbf{x})$. 
The forward process, which systematically degrades the underlying structure of the data manifold, can be formalized as an Itô Stochastic Differential Equation (SDE):
$$d\mathbf{x} = f(t)\mathbf{x} \, dt + g(t) \, d\mathbf{w}$$
where $f(t)$ represents the deterministic drift coefficient, $g(t)$ is the scalar diffusion coefficient controlling the injection rate of stochastic noise, and $\mathbf{w}$ is a standard Brownian motion (Wiener process).

A foundational result in stochastic calculus established by Anderson~\cite{and82} dictates that the reverse-time process—moving from the unstructured prior $p_{\text{prior}}$ back to the data manifold $p_{\text{data}}$—is also a diffusion process running backward in time. It satisfies the reverse-time SDE:
$$d\mathbf{x} = \left[ f(t)\mathbf{x} - g(t)^2 \nabla_\mathbf{x} \log p_t(\mathbf{x}) \right] dt + g(t) \, d\bar{\mathbf{w}}$$
where $d\bar{\mathbf{w}}$ is a backward Wiener process running from $t=T$ to $t=0$, and $p_t(\mathbf{x})$ represents the marginal probability density of $\mathbf{x}$ at time $t$. The term $\nabla_\mathbf{x} \log p_t(\mathbf{x})$ is the Stein score function—a time-dependent vector field pointing toward regions of higher probability density at noise scale $t$.

\subsubsection*{The Probability Flow ODE}
A critical insight established by Song et al. \cite{sony21} is that for every forward~SDE, there exists a deterministic counterpart known as the Probability Flow Ordinary Differential Equation (PF-ODE). The PF-ODE is formulated as:
$$\frac{d\mathbf{x}}{dt} = f(t)\mathbf{x} - \frac{1}{2}g(t)^2 \nabla_\mathbf{x} \log p_t(\mathbf{x})$$
The SDE and the PF-ODE are sharing the exact same marginal probability densities $\{p_t(\mathbf{x})\}_{t=0}^T$ at every instantaneous time step. Consequently, generating a sample can be executed either stochastically (via the reverse SDE) or deterministically (by integrating the PF-ODE). In both instances, the sole unknown quantity required to parameterize the generative trajectory is the score function $\nabla_\mathbf{x} \log p_t(\mathbf{x})$.
\vspace{-8pt}

\subsection*{Score-Based Diffusion: Unifying DDPMs and SGMs}
DDPMs and SGMs were developed independently using discrete-time formulations. The~continuous-time SDE framework unifies them as specific parameterizations of the scalar drift $f(t)$ and diffusion $g(t)$ coefficients. However, understanding the operational mechanics of this unification requires grounding the~framework in the explicit variational lower bound (ELBO) established by Ho et al. \cite{ho20}.

\subsubsection*{The Discrete ELBO Starting Point in Ho's Formulation}
In the discrete-time formulation of DDPM, the forward corruption process is defined as a non-learnable Markov chain that parameterizes a latent variable hierarchy $\mathbf{x}_1, \mathbf{x}_2, \dots, \mathbf{x}_T$ conditioned on the data $\mathbf{x}_0 \sim p_{\text{data}}$:
$$q(\mathbf{x}_{1:T} | \mathbf{x}_0) = \prod_{t=1}^T q(\mathbf{x}_t | \mathbf{x}_{t-1}), \quad q(\mathbf{x}_t | \mathbf{x}_{t-1}) = \mathcal{N}(\mathbf{x}_t; \sqrt{1-\beta_t}\mathbf{x}_{t-1}, \beta_t \mathbf{I})$$
where $\beta_t \in (0, 1)$ is a predefined variance schedule. The generative reverse process is modeled as another Markov chain starting from a standard Gaussian prior $p(\mathbf{x}_T) = \mathcal{N}(0, \mathbf{I})$:
$$p_\theta(\mathbf{x}_{0:T}) = p(\mathbf{x}_T) \prod_{t=1}^T p_\theta(\mathbf{x}_{t-1} | \mathbf{x}_t), \quad p_\theta(\mathbf{x}_{t-1} | \mathbf{x}_t) = \mathcal{N}(\mathbf{x}_{t-1}; \mathbf{\mu}_\theta(\mathbf{x}_t, t), \mathbf{\Sigma}_\theta(\mathbf{x}_t, t))$$
To train the parameterized transition kernels $p_\theta(\mathbf{x}_{t-1}|\mathbf{x}_t)$, Ho et al. \cite{ho20} invoke the variational principle to minimize the negative log-likelihood $-\log p_\theta(\mathbf{x}_0)$ via its variational upper bound 
(the negative ELBO):
$$\mathcal{L}_{\text{VBound}} = \mathbb{E}_q \left[ \log \frac{q(\mathbf{x}_{1:T} | \mathbf{x}_0)}{p_\theta(\mathbf{x}_{0:T})} \right] \ge -\mathbb{E} \left[ \log p_\theta(\mathbf{x}_0) \right]$$
By applying Bayes' rule to expand the forward joint transitions as a function of the conditional posteriors $q(\mathbf{x}_{t-1}|\mathbf{x}_t, \mathbf{x}_0)$, Ho et al. \cite{ho20} decompose this objective into a sum of localized KL divergence terms:
$$\mathcal{L}_{\text{VBound}} = \mathbb{E}_q \Big[ \underbrace{D_{\text{KL}}\big(q(\mathbf{x}_T | \mathbf{x}_0) \parallel p(\mathbf{x}_T)\big)}_{L_T} + \sum_{t=2}^T \underbrace{D_{\text{KL}}\big(q(\mathbf{x}_{t-1} | \mathbf{x}_t, \mathbf{x}_0) \parallel p_\theta(\mathbf{x}_{t-1} | \mathbf{x}_t)\big)}_{L_{t-1}} - \underbrace{\log p_\theta(\mathbf{x}_0 | \mathbf{x}_1)}_{L_0} \Big]$$
where $L_T$ contains no learnable parameters and measures how closely the final step of the noise schedule matches the standard Gaussian prior, $L_0$ is a reconstruction term modeling the final projection from the~pixel-space latent $\mathbf{x}_1$ back to the clean image $\mathbf{x}_0$, and $L_{t-1}$ represents the core optimization target for all intermediate layers, enforcing that the learned reverse step matches the true, tractably analytical forward posterior step $q(\mathbf{x}_{t-1} | \mathbf{x}_t, \mathbf{x}_0)$. Because the forward process is Gaussian, the true posterior conditioned on the clean image $\mathbf{x}_0$ is analytically tractable: \vspace{-5pt}
$$q(\mathbf{x}_{t-1} | \mathbf{x}_t, \mathbf{x}_0) = \mathcal{N}\big(\mathbf{x}_{t-1}; \, \tilde{\mu}_t(\mathbf{x}_t, \mathbf{x}_0), \, \tilde{\beta}_t \mathbf{I}\big) \vspace{-3pt} $$  
where the true posterior mean is a weighted combination of the noisy state $\mathbf{x}_t$ and the original target~$\mathbf{x}_0$: 
$$\tilde{\mu}_t(\mathbf{x}_t, \mathbf{x}_0) = \frac{\sqrt{\alpha_t}(1-\bar{\alpha}_{t-1})}{1-\bar{\alpha}_t}\mathbf{x}_t + \frac{\sqrt{\bar{\alpha}_{t-1}}\beta_t}{1-\bar{\alpha}_t}\mathbf{x}_0$$
with $\alpha_t = 1 - \beta_t$ and $\bar{\alpha}_t = \prod_{s=1}^t \alpha_s$.

While DDPMs produce high-quality images, they required thousands of discrete steps for sampling. Song {\it et al.} \cite{sonj21} introduced a non-Markovian DDIM process that allows faster sampling without retraining. 
\vspace{-16pt}

\subsubsection*{Transition to Continuous-Time Score Matching}
The connection between this discrete variational~formulation and continuous-time score vector fields emerges when taking the limit $T \to \infty$. As the time~steps form a continuum, the forward Markovian corruption converges to the Variance Preserving (VP) SDE:
$$d\mathbf{x} = -\frac{1}{2}\beta(t)\mathbf{x} \, dt + \sqrt{\beta(t)} \, d\mathbf{w}$$
where $\beta(t)$ is the continuous equivalent of the discrete noise schedule. Under this continuous formulation, the marginal perturbation kernel simplifies to $q(\mathbf{x}_t | \mathbf{x}_0) = \mathcal{N}(\mathbf{x}_t; \sqrt{\bar{\alpha}_t}\mathbf{x}_0, (1-\bar{\alpha}_t)\mathbf{I})$.
Concurrently, classical SGMs (Song and Ermon \cite{sony19}) perturb data by adding progressively larger volumes of noise without scaling the data vector, mapping instead to the Variance Exploding (VE) SDE:
$$d\mathbf{x} = \sqrt{\frac{d\sigma^2(t)}{dt}} \, d\mathbf{w}$$
where $\sigma(t)$ is a monotonically increasing noise schedule, yielding the kernel $q(\mathbf{x}_t | \mathbf{x}_0) = \mathcal{N}(\mathbf{x}_t; \mathbf{x}_0, \sigma^2(t)\mathbf{I})$.

To optimize across both configurations without managing the intractable true marginal density $\nabla_\mathbf{x} \log p_t(\mathbf{x})$, we rely on the foundational equivalence established by Vincent \cite{vin11}. By introducing Denoising Score Matching (DSM), Vincent demonstrated that one can bypass the marginal density entirely by optimizing directly against the analytical score of the conditional perturbation kernel $\nabla_{\mathbf{x}_t} \log q(\mathbf{x}_t | \mathbf{x}_0)$
$$\mathcal{L}_{\text{DSM}}(\theta) = \mathbb{E}_{t \sim \mathcal{U}[0,T]} \mathbb{E}_{\mathbf{x}_0 \sim p_{\text{data}}} \mathbb{E}_{\mathbf{x}_t \sim q(\mathbf{x}_t|\mathbf{x}_0)} \left[ \lambda(t) \left\| s_\theta(\mathbf{x}_t, t) - \nabla_{\mathbf{x}_t} \log q(\mathbf{x}_t | \mathbf{x}_0) \right\|_2^2 \right]$$
For a Gaussian perturbation kernel, this target score evaluates to $\nabla_{\mathbf{x}_t} \log q(\mathbf{x}_t | \mathbf{x}_0) = -\frac{\mathbf{x}_t - \mathbf{\mu}_t}{\sigma_t^2} = -\frac{\epsilon}{\sigma_t}$, where $\mathbf{\epsilon} \sim \mathcal{N}(0, \mathbf{I})$ is the actual noise vector injected during the forward pass.

The profound innovation in Ho's framework was utilizing the $\mathbf{x}_0$ reparameterization trick—expressing $\mathbf{x}_0$ strictly as a function of the visible noisy sample $\mathbf{x}_t$ and the latent noise $\epsilon$:
$$\mathbf{x}_0 = \frac{1}{\sqrt{\bar{\alpha}_t}}\left(\mathbf{x}_t - \sqrt{1-\bar{\alpha}_t}\epsilon\right)$$
Substituting this definition of $\mathbf{x}_0$ back into the analytical formula for the true posterior mean $\tilde{\mu}_t(\mathbf{x}_t, \mathbf{x}_0)$ allows the network's parameterization to mimic the exact algebraic structure of the true mean:
$$\mu_\theta(\mathbf{x}_t, t) = \frac{1}{\sqrt{\alpha_t}}\left(\mathbf{x}_t - \frac{\beta_t}{\sqrt{1-\bar{\alpha}_t}}\epsilon_\theta(\mathbf{x}_t, t)\right)$$
When these formulations are substituted back into the discrete intermediate ELBO term ($L_{t-1}$), the~terms $x_t$, $\alpha_t$, and $\beta_t$ factor out. The heavy, path-dependent variational lower bound collapses elegantly into a simple, single-step noise estimation loss:
$$\mathcal{L}_{\text{simple}}(\theta) = \mathbb{E}_{t, \mathbf{x}_0, \epsilon} \left[ \left\| \epsilon - \epsilon_\theta(\mathbf{x}_t, t) \right\|_2^2 \right]$$
This collapse unifies the optimization profiles of variational diffusion models and score-based models, exposing the explicit structural identity linking the noise predictor to the continuous Stein score function:$$s_\theta(\mathbf{x}_t, t) \equiv -\frac{\epsilon_\theta(\mathbf{x}_t, t)}{\sigma_t}$$

\subsection*{Flow Matching: Moving Beyond Stochastic Trajectories}
While score-based diffusion models provide a rigorous framework, their trajectories are inherently anchored to stochastic processes. This constraints the paths to be highly non-linear, which increases numerical complexity during ODE integration and necessitates a large number of sampling steps.

Flow Matching (FM) (Albergo and Vanden-Eijnden \cite{alb23}, Lipman et al. \cite{lip23}) circumvents this limitation by abandoning the underlying SDE framework entirely. Instead, FM models generative trajectories strictly through a deterministic fluid-dynamics perspective using a time-dependent velocity vector field $\mathbf{v}_t(x)$. The~generative process is governed by a simple, non-stochastic ODE:
$$\frac{d\mathbf{x}}{dt} = \mathbf{v}_t(x)$$

\subsubsection*{Conditional Flow Matching (CFM)}
The objective of FM is to train a neural network $\mathbf{v}_\theta(\mathbf{x}, t)$ to regress the target velocity field $\mathbf{v}_t(\mathbf{x})$ that generates the path from noise to data. Because the~true marginal velocity field $\mathbf{v}_t(\mathbf{x})$ is mathematically intractable, FM introduces a formulation analogous to DSM, termed Conditional Flow Matching (CFM).
By conditioning the trajectory on the boundary states (typically $\mathbf{x}_0 \sim p_{\text{data}}$ and $\mathbf{x}_1 \sim p_{\text{prior}}$, setting $T=1$), we define a conditional probability path $p_t(\mathbf{x}|\mathbf{x}_0, \mathbf{x}_1)$ and its corresponding conditional velocity field $\mathbf{v}_t(\mathbf{x}|\mathbf{x}_0, \mathbf{x}_1)$. The objective function is formulated as:
$$\mathcal{L}_{\text{CFM}}(\theta) = \mathbb{E}_{t \sim \mathcal{U}[0,1]} \mathbb{E}_{\mathbf{x}_0 \sim p_{\text{data}}} \mathbb{E}_{\mathbf{x}_1 \sim p_{\text{prior}}} \mathbb{E}_{\mathbf{x} \sim p_t(\mathbf{x}|\mathbf{x}_0, \mathbf{x}_1)} \left[ \left\| \mathbf{v}_\theta(\mathbf{x}, t) - \mathbf{v}_t(\mathbf{x} | \mathbf{x}_0, \mathbf{x}_1) \right\|_2^2 \right]$$ 

\subsubsection*{Optimal Transport and Rectified Flows}
The flexibility of CFM lies in the fact that the conditional paths do not need to be derived from a stochastic diffusion process. By choosing a Gaussian conditional path that interpolates linearly between data and noise: $p_t(\mathbf{x} | \mathbf{x}_0, \mathbf{x}_1) = \mathcal{N}\left(\mathbf{x}; \, (1-t)\mathbf{x}_0 + t\mathbf{x}_1, \, \sigma_0^2 \mathbf{I}\right)$, the~exact conditional velocity field reduces to a constant straight line:
$$\mathbf{v}_t(\mathbf{x} | \mathbf{x}_0, \mathbf{x}_1) = \frac{d}{dt}\left[(1-t)\mathbf{x}_0 + t\mathbf{x}_1\right] = \mathbf{x}_1 - \mathbf{x}_0$$
This specific formulation—often referred to as Optimal Transport FM or Rectified Flow (Liu et al.~\cite{liu23}) straightens the generative trajectories. By minimizing the curvature of the integration paths, FM models dramatically reduce the number of function evaluations (NFE) required during inference, allowing high-fidelity image synthesis in as few as 4 to 10 deterministic steps.
\vspace{-10pt}

\subsection*{The Unified Mathematical Synthesis}
The profound realization linking these three frameworks is that the Probability Flow ODE of score-based diffusion is identical to a Flow Matching ODE. By equating the deterministic velocity field $\mathbf{v}_t(\mathbf{x})$ directly to the drift and score-diffusion terms of the PF-ODE, we obtain the universal trajectory identity:
$$\mathbf{v}_t(\mathbf{x}) \equiv \underbrace{f(t)\mathbf{x}}_{\text{Drift Field}} - \underbrace{\frac{1}{2}g(t)^2 \nabla_\mathbf{x} \log p_t(\mathbf{x})}_{\text{Score-Driven Diffusion Field}}$$

Through this synthesis, the evolution of trajectory-based models can be viewed as a continuous optimization of path geometry: transitioning from stochastic paths (SDEs), to deterministic curves (Scores and PF-ODEs), and ultimately to straight vector lines (Flow Matching), systematically refining sampling efficiency while preserving the expressive capacity of the underlying data manifold.
\vspace{-20pt}

\subsection*{Image-to-Image Schrödinger Bridge: Entropic Optimal Transport} \vspace{-3pt}
Image-to-Image Schrödinger Bridge (SB) \cite{lig23} generalizes the Flow Matching (FM) framework by formulating image generation as an Entropic Optimal Transport (EOT) problem that brings back the~benefits of stochasticity. Technically, FM is the "zero-noise limit" of the~SB, which recovers the optimal path that minimizes the KL divergence to a reference prior SDE, while satisfying exactly the marginal distributions at both the initial and final times. In image modeling, 
SBs can seamlessly construct optimal paths between domains, such as corrupted images and clean images, or sketches and photorealistic outputs.
\vspace{-8pt}

\subsection*{Latent Diffusion Models} \vspace{-4pt}
In order to generate a $1024 \times 1024$ image, denoising diffusion models have to calculate over a million pixels at every timestep. This is computationally expensive. To handle high-resolution images efficiently, Rombach {\it et al.} \cite{rom22} apply 
the~diffusion process in the latent space of a pre-trained autoencoder. 

A Latent Diffusion Model (LDM) consists of three blocks:
(i) The VAE: An encoder that turns a high-resolution image into a compressed latent representation ($z$). VAEs are often labeled as f8 or f16, which refers to the downsampling factor: e.g., for $512\times512$ input image, f8 refers to $512 \div 8$, i.e.,~$64\times64$ latents, which was used for Stable Diffusion v1 and v2.
f4 provides more detail but is much slower.
f16/f32 are very fast, but the VAE "forgets" details. 
(ii) Diffusion Process: The U-Net has to denoise the smaller latent grid (e.g., $64 \times 64$).
(iii) The Decoder: Once the~latents are "clean," the VAE decoder expands them back up into a full-resolution, high-fidelity image. 

To ensure the latent space is well-behaved for a diffusion model, Rombach {\it et al.} \cite{rom22} explored two types of regularization to train the VAE: 
(i) Kullback-Leibler: Similar to a standard VAE, it imposes a penalty on the~learned latents to keep them close to a standard normal distribution. 
(ii) Vector Quantization: It uses a codebook layer (like VQ-VAE) but the quantization is often integrated into the decoder. This makes the latent space discrete.
Stable Diffusion primarily uses the KL-regularized VAE because it results in smoother, continuous latent spaces that are easier for the Diffusion U-Net to model.
Esser {\it et al.} \cite{ess21} developed the VQ-GAN (Vector Quantized GAN), which provided perceptual compression necessary~to represent images as compact codes for transformer based  denoising while maintaining sharp visual details.
 
During the training of the Diffusion Model (the U-Net), the~VAE is usually frozen. We only use it to encode images into latents once, and then the U-Net 
is trained to learn how to denoise those latents.

In the evolution of image modeling, LDM \cite{rom22} serves as a bridge between diffusion models and  Foundation Image Models. The significance of LDM lies in: (i) Efficiency: Reducing the search space from $H \times W \times 3$ to a latent representation (typically 1/8th the size) allowed researchers to train models on billions of images. (ii) Conditioning: It introduced Cross-Attention layers into the diffusion backbone, providing a flexible way to inject text prompts, depth maps, or semantic masks into the modeling process.

\section*{Foundation Image Models: The Universal Image Prior}
Foundation Image Models (FIM) learn a universal representation of the natural image manifold by scaling parameters into the billions and learning across massive cross-modal datasets. This transition from Deep Generative Models (DGM) to FIMs shifts the paradigm from strict, single-task generative generation to versatile, zero-shot open-domain understanding and synthesis.
\vspace{-10pt}

\subsection*{What Distinguishes Foundation Image Models from Deep Generative Models}
While the U-Net has been the workhorse of original LDMs, the transition to FIMs was triggered by the realization that Transformers, which revolutionized natural language processing (NLP), could be tamed for images. By replacing the U-Net with a Diffusion Transformer (DiT) architecture, proposed by Peebles and Xie \cite{pee23}, see Figure 3, researchers could finally apply Scaling Laws to image processing and vision.
DGMs leverage convolutional frameworks (like U-Nets) which carry a strong structural inductive bias assuming localized pixel correlations. FIMs employ patch-based DiTs, utilizing global self-attention to capture long-range dependencies.
By treating the latent representations of an image as a sequence of tokens ("Patchification")—analogous to words in a sentence—DiTs apply the same highly scalable architectures that power Large Language Models (LLMs) to model the continuous visual world. Furthermore, AR foundation models, which predict the~next image token in a quantized latent space offer unified frameworks for both text and image generation. However, a model does not become a "Foundation Model" just because it uses a Transformer backbone. Other factors distinguishing DGMs from FIMs are:

Parameter and Dataset Scale: DGMs are typically smaller networks trained on specialized, narrow datasets (e.g., human face sets). FIMs scale to billions of parameters and undergo pre-training on diverse, web-scale datasets spanning billions of image-text pairs.

\begin{figure}[t!]
\centering
\includegraphics[width=0.9\textwidth]{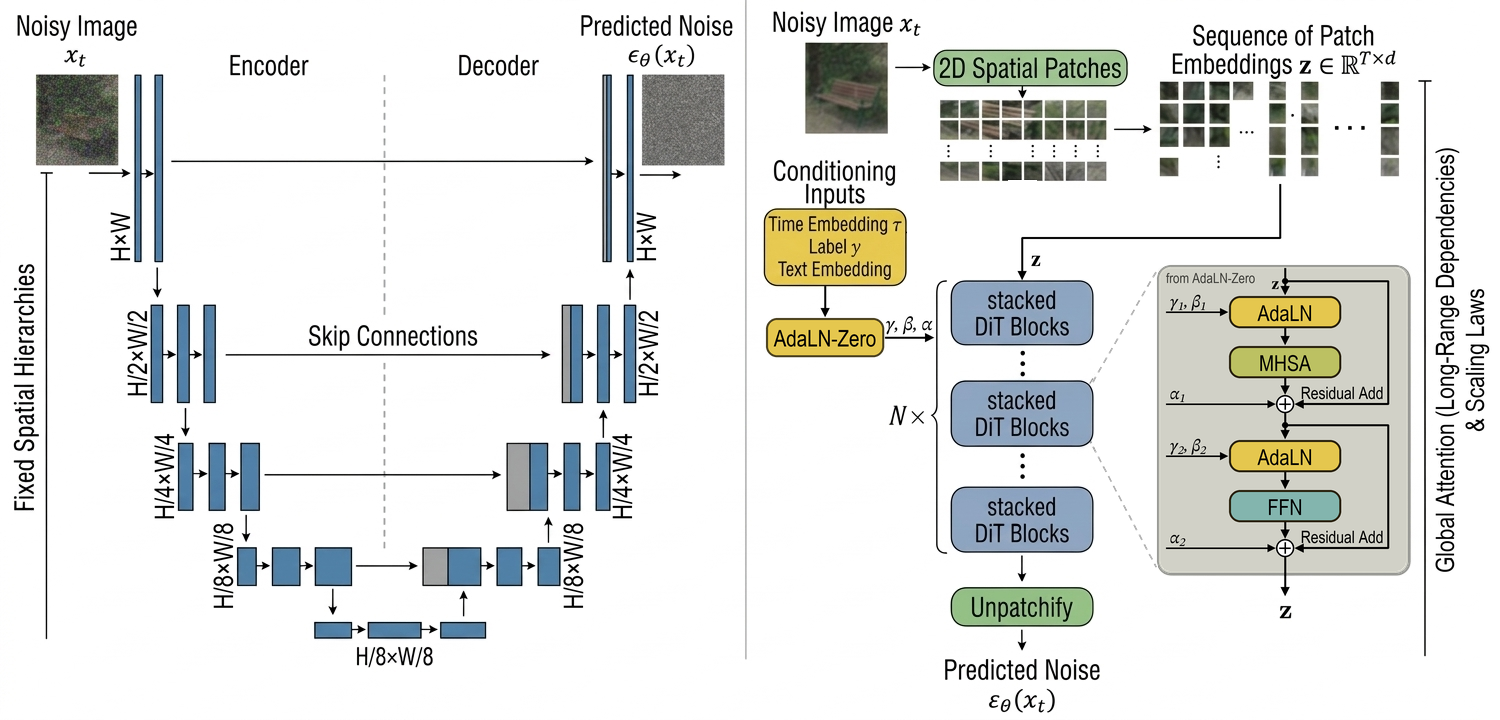} \vspace{-12pt} \\
(a) \hspace{5cm} (b) \vspace{-18pt}
\caption{The architecture of Foundation Image Models: Transition from Inductive Bias of CNNs to Scaling Law of transformers. (a) Traditional diffusion backbone: Convolutional U-Net, (b) The foundation model backbone: Diffusion Transformer~(DiT)~\cite{pee23}.
Illustration is generated using Google Gemini based on author prompts for layout design. The content is verified by the author.} 
\vspace{-10pt} \label{fig:sonn-sr}
\end{figure}

Multiple Modalities: The integration of multiple modalities (text, images, video, and sequence semantics) within shared latent and embedding spaces is a fundamental requirement of FIM. Traditional DGMs are largely unimodal; they focus strictly on mapping noise vectors to pixel spaces based on visual statistics. FIMs, however, use joint cross-modal alignment mechanisms (such as CLIP) to structurally intertwine textual semantics with high-dimensional visual features.
This multimodal alignment 
guides the model's reverse sampling mechanics directly to a highly specific, semantically correct sub-manifold, eliminating the~structural "hallucinations" common in older DGM configurations. 

Task Generalization (Narrow vs. Zero-Shot): DGMs are bound to narrow task scopes and struggle with out-of-distribution inputs. FIMs exhibit robust zero-shot generalization, enabling them to successfully execute 
tasks for which they were never explicitly or narrowly trained.
In terms of versatility as an image prior, DGMs are fragile priors; if deployed to solve an inverse problem outside their training distribution, they can force incorrect features into the reconstruction. FIMs act as zero-shot universal image priors that can be plugged into arbitrary iterative solvers to fix any 
degradation without retraining the base network.
\vspace{-30pt}

\subsection*{Classification of Foundation Image Models}
\vspace{-3pt}

Foundation image models (FIM) can be broadly classified into the following categories: 

\subsubsection*{Representation Models (Vision-language alignment)}
Multimodal representation models are the bridge between language and vision and often serve as the~backbone for generative and discriminative FIM.
Contrastive language-image pre-training (CLIP) by~OpenAI and Sigmoid Loss for Language Image pre-training (SigLIP) by Google aim to map images and text into the same high-dimensional space. 
DIstillation with NO labels (DINOv2) by Meta leverages self-supervised learning on massive image datasets to learn deep visual features purely from pixels without relying on human-annotated text or labels.

\subsubsection*{Promptable Generative FIMs (Synthesizing pixels)} 
DALL-E 2 by OpenAI and Imagen by Google are among the pioneers based on the U-net backbone.
Stable Diffusion XL (SDXL)~\cite{pod24} features improved LDM architecture, including a larger U-Net backbone and addition of a refinement stage. It is significant for introducing micro-conditioning (conditioning on image size and cropping parameters) and a two-stage pipeline (base model + refiner) to improve visual fidelity. 
Stable Diffusion 3.0 (SD3) \cite{ess24} is significant for moving away from the U-Net entirely in favor of a Transformer-based backbone (building on the DiT architecture by Peebles and Xie \cite{pee23}) and utilizing Flow Matching (Rectified Flow).


Stable Diffusion 3.5 \cite{ess24} and FLUX.1, the state-of-the-art open-weights models, represent the critical architectural shift from Convolutional U-Nets to DiT architecture and the mathematical shift from standard diffusion to Flow Matching (Rectified Flows).
These models are the real-world culmination of LDM and Schrödinger Bridge concepts discussed earlier. By replacing the U-Net with a scalable Transformer backbone, they treat latent image patches as sequences, aligning image generation with the scaling laws of  LLMs. Furthermore, they utilize Flow Matching (which learns straight-line trajectories between noise and data), drastically reducing sampling steps and improving text-to-image semantic alignment. 

\subsubsection*{Promptable Discriminative Foundation Vision Models (Parsing and segmenting)}
Segment Anything Model (SAM) 2 defines the "promptable" era for computer vision tasks, where~foundation models not only generate images, but also solve zero-shot segmentation and tracking tasks.
SAM~acts as a universal, zero-shot segmentation prior. A user can "prompt" the model with a click, a bounding box, or text.
\vspace{-9pt}

\subsection*{Solving Inverse Problems in Imaging Using FIM Priors}  \vspace{-4pt}

We can classify methods to solving inverse problems using FIM priors as training-time, supervised finetuning approaches vs. inference-time zero-shot physics-guided optimization on frozen FIM manifolds.

\subsubsection*{A. Training-time Task-Specific, Supervised Fine-Tuning of FIMs via Conditional Adapters} 
A lightweight auxiliary module (like ControlNet or an adapter) is added to a large pretrained FIM 
to take a low-quality/degraded image $\mathbf{y}$ as input and inject its features directly into the frozen FIM backbone's intermediate layers.
The modified FIM learns the posterior $p(\mathbf{x}\mid \mathbf{y})$ implicitly by training the added layers (backbone is frozen) on paired low-quality and high-quality data. Inference is single pass. 
This approach can handle blind/unknown degradations (real-world blur, noise) because it does not require an explicit, analytically defined degradation operator $\mathbf{A}$. However, it is task-specific, a model trained for $\times4$~SR cannot perform inpainting or deblurring without being retrained.
Popular task-specific inverse problem solver frameworks using FIMs include StableSR~\cite{wan24}, its evolution DiffBIR (Diffusion Blind Image Restoration) \cite{lin24}, ControlNet, SR3, ResShift, DiT4SR, SD3-ControlNet, and FluxSR.

\noindent Conditional Injection: U-Net vs. DiT Backbone:
StableSR~\cite{wan24} was designed around a U-Net architecture relying on the generative priors of Stable Diffusion v1.5 and v2.1 combined with a ControlNet-style time-aware encoder. 
StableSR injects the conditional LR image into a frozen Stable Diffusion U-Net by utilizing an auxiliary network called a Time-Aware Encoder paired with Spatial Feature Transform (SFT) layers.
The~SFT layers are inserted right after the Residual blocks in both the downsampling and upsampling blocks of the frozen U-Net. The injection works through affine transformations (scaling and shifting):
$\text{SFT}(F_{\text{unet}}) = \gamma \odot F_{\text{unet}} + \beta$,
where $\odot$ represents element-wise multiplication.

Recently, the community has transitioned from U-Net to Diffusion Transformer (DiT) and Multimodal DiT (MM-DiT) backbones. The core philosophy of conditional injection remained the same, but U-Net feature injection has been replaced with spatial token interaction inside cross-attention blocks:

1. DiT4SR (Taming Diffusion Transformer for Real-World SR) adapts Stable Diffusion 3.5 (which uses an MM-DiT architecture) for real-world super-resolution.
Because transformers handle inputs as tokens, DiT4SR discards the external ControlNet pipeline. Instead, it tokenizes the LR image and directly integrates the LR stream into the original DiT blocks via attention mechanisms.

2. SD3-ControlNet for Super-Resolution:
Similar to StableSR's method of anchoring a frozen text-to-image model, SD3-ControlNet freezes the core MM-DiT layers and trains a copy of the DiT blocks to ingest LR visual tokens, providing the necessary guidance to recover fine high-frequency details.

3. FluxSR and StrSR (Flow-Based DiT SR):
Recent task-specific SR models like FluxSR and StrSR (Spectral and Trajectory Regularization for DiT SR) have ported 
generative SR to flow-matching transformers, such as FLUX that use DiT backbones based on Rectified Flow (straight-line trajectories).
They focus on addressing DiT-specific artifacts (like high-frequency grid patterns caused by spectral leakage) while distilling the massive transformer for fast, few-step generation.

\subsubsection*{B. Inference-Time Zero-Shot, Physics-Guided Optimization on Frozen FIM Manifolds}
We use a known degradation operator ($A$) to guide a frozen, unconditional FIM during inference. There are 3 paradigms:

Paradigm 1:
The Score-Based Posterior Sampling 

An unconditional FIM generates an image by calculating the unconditional score function, $\nabla_{\mathbf{x}_t} \log p(\mathbf{x}_t)$, which is a vector field pointing toward the closest "natural image" manifold.
To solve an inverse problem, we don't want any natural image; we want a natural image that matches our low-quality measurement~$\mathbf{y}$. By applying Bayes' rule and taking the gradient, the conditional score $\nabla_{\mathbf{x}_t} \log p(\mathbf{x}_t \mid \mathbf{y})$ can be split into: 
$$\underbrace{\nabla_{\mathbf{x}_t} \log p(\mathbf{x}_t \mid \mathbf{y})}_{\text{Conditional Score (Where to go)}} = \underbrace{\nabla_{\mathbf{x}_t} \log p(\mathbf{x}_t)}_{\text{Prior Score (From Frozen Model)}} + \underbrace{\nabla_{\mathbf{x}_t} \log p(\mathbf{y} \mid \mathbf{x}_t)}_{\text{Likelihood Score (Physical Guidance)}}$$

Every single timestep $t$ during the diffusion reverse (sampling) process:
1) The Prior Score (Model): Pushes the current noisy image $\mathbf{x}_t$ toward looking like a sharp, realistic, natural image.
2) The Likelihood Score (Physics): Measures how badly the current image violates the physical measurement $\mathbf{y}$, and calculates a mathematical correction gradient to pull it back into alignment.
Examples include DPS (Diffusion Posterior Sampling), DDNM, MCG (Manifold Constrained Guidance), PiGDM.

Score-based posterior sampling is a universal inverse problem solver. The same model weights can solve different inverse problems by swapping $\mathbf{A}$ at runtime 
without ever needing retraining. However, the approach is non-blind. To compute the likelihood score, we must know the exact definition of $\mathbf{A}$. Furthermore, because we have to compute a gradient through the forward operator at every single step of the trajectory, inference is slow—often requiring hundreds of full backpropagation passes per image.

\newpage

Paradigm 2: Latent Space / Manifold Optimization

Latent space/manifold optimization is a natural evolution of the Deep Image Prior (DIP) approach. 
In~2017, Ulyanov et al. \cite{uly18} discovered an interesting property of convolutional neural network architectures: a network's structure itself acts as an image prior, even before it is trained on anything. To solve an inverse problem using DIP, we take an untrained, randomly initialized generator network $G_\theta$, feed it with random noise $\mathbf{z}$, and optimize the network weights $\theta$ to match the degraded image $\mathbf{y}$:
$$\min_{\theta} \| \mathbf{A}(G_\theta(\mathbf{z})) - \mathbf{y} \|_2^2$$
\vspace{-20pt}

While groundbreaking, DIP has two major weaknesses that made it unreliable:
1) Overfitting to Noise: If we let the optimization run too long, the network will overfit the degraded input image.
Hence, it requires careful "early stopping" before the model starts to memorize the degradation/artifacts.
2)~No High-Level Semantics: Because the network is untrained, it only provides inductive bias
(smoothness, local gradients). 
It doesn't know semantic concepts like "human eye" or "brick texture."

Latent Space Optimization (often referred to in GAN literature as GAN Inversion or Deep Generative Prior (DGP)) was developed to fix these limitations.
Instead of an untrained network, we use a pre-trained FIM (like StyleGAN or a Latent Diffusion Model) that has already spent significant amount of compute learning the high-level semantic features of natural images.
Because the model weights $\theta$ are frozen, FIM defines a rigid, low-dimensional Latent Manifold. We shift the optimization from a fragile, high-dimensional weight space into a highly constrained, semantically rich low-dimensional latent code~($\mathbf{z}$):
$$\min_{\mathbf{z}} \| \mathbf{A}(G_{\text{frozen}}(\mathbf{z})) - \mathbf{y} \|_2^2$$
That is, we search the fixed, low-dimensional latent space ($\mathbf{z}$) of a frozen generator via gradient descent to find the specific latent code that minimizes $\|\mathbf{A}(G(\mathbf{z})) - \mathbf{y}\|_2^2$. 
Examples include PULSE, GAN Inversion.

Paradigm 3: Optimization via Variable Splitting (Plug-and-Play)

This is a direct extension of the PnP approach (discussed earlier), which decouples the physical data-fidelity step from the regularizer (prior) step using splitting algorithms, discussed earlier in the paper.
Diffusion-based FIM (e.g., Stable Diffusion, FLUX) are, at their core, highly advanced denoisers. They were trained explicitly on billions of images to take a noisy input and predict the clean underlying image. To use the FIM as a modular PnP block, we simply pass the intermediate image $\mathbf{x}^{(k+1)}$ into the FIM's core denoising block (the U-Net or DiT) for a single-step or few-step denoiser pass. The FIM uses its rich semantic prior to project the image back toward the natural image manifold. Examples include PnP-ADMM, PnP-HQS.

\section*{Implicit Neural Representation}
\label{implicit}

The shift from pixel-based discrete representation to continuous function representation is a mathematically elegant transformation in image processing. 
An Implicit Neural Representation (INR) defines an image as a continuous, differentiable function parameterized by a neural network, $f_\theta(x,y) \rightarrow \mathbf{v}$,
where the inputs $(x,y)$ are continuous spatial coordinates, the output $\mathbf{v}$ is the RGB color at that coordinate. 
The~network $f_\theta$ is a Multilayer Perceptron (MLP) that learns the mapping.
To render an image, we simply query the network at every $(x,y)$ of the desired resolution, e.g., $f_\theta (0.5,0.1)$ returns an RGB~value. This~means the underlying representation has infinite resolution—we can zoom in indefinitely.

The true power of INRs lies in the fact that the mapping $f_\theta$ is a fully differentiable computational graph. This grants us two distinct types of gradients, each unlocking different capabilities: \\
A. Gradients with respect to Coordinates ($\nabla_\mathbf{x} f$):
Because the input coordinates are continuous variables, we can calculate the derivative of the image with respect to spatial position.
We don't need edge detection filters (like Sobel operators).
We can calculate the exact mathematical gradient of the image intensity.
\\
B. Gradients with respect to Weights ($\nabla_\theta f$):
This enables learning. By backpropagating errors from a target (e.g., a blurry photo) into the weights $\theta$, the network can match the observations. 

\subsubsection*{INR as an Architectural Prior}
When an INR is initialized with random weights, it doesn't produce white noise. Surprisingly, it produces smooth, coherent, low-frequency patterns. That is, the structure of a neural network (convolutions or fully connected layers) is naturally biased toward generating smooth images rather than high-frequency noise. This is closely related to the Deep Image Prior (DIP) concept, where the structure of the network itself regularizes the signal.
If we train an INR to replicate a noisy image, the network will learn the "clean" image structure first (because it's easy to represent) and will struggle to learn the random noise (because it's hard to represent).
This makes INRs powerful unsupervised tools for tasks like Denoising (stop training before it learns the noise).

\subsubsection*{Overcoming the "Spectral Bias"}
A major challenge with standard MLPs is that they are notoriously bad at learning high-frequency details (sharp edges and fine textures). This is known as Spectral Bias—networks converge on low-frequency components much faster.
To make INRs viable for high-fidelity imaging, Sitzmann {\it et al.} \cite{sit20} introduced Positional Encoding (similar to Transformers) or sinusoidal activation functions (SIREN):
$$\gamma(p) = (\sin(2^0 \pi p), \cos(2^0 \pi p), \dots, \sin(2^L \pi p), \cos(2^L \pi p))$$
By mapping coordinates into a higher-dimensional Fourier feature space, the network can tune specific frequencies, allowing it to capture both the smooth gradient of sky and the sharp edge of a building.

\section*{Discussion and Conclusions}
Early handcrafted models
provided rigorous mathematical guarantees and interpretability but struggled to capture the structure and texture of natural images. The deep learning era, spearheaded by CNNs and generative models~\cite{tek22}, shifted the focus toward learning representations from large datasets. 
Within the family of diffusion models, which represent the state of the art: DDPM/SMLD learn to reverse a curved, stochastic path of noise. SDE/ODE unified them into a single continuous formulation. Score Matching learns the ODE velocity indirectly via the score. FM learns the ODE velocity directly and straightened the path to make it faster.
LDMs made high-quality image generation possible on consumer GPUs.

Today, we have entered the era of \textit{Foundation Image Models}. By replacing the U-Net with a Diffusion Transformer (DiT), we can apply Scaling Laws to image processing and vision tasks, developing models that demonstrate zero-shot capabilities—the ability to perform tasks they aren't explicitly trained for. Architectures built upon Vision Transformers (ViTs) and 
Latent Diffusion Models (LDMs) are no longer confined to narrow tasks. Models such as Stable Diffusion, Segment Anything Model (SAM), and large Vision-Language Models (VLMs) demonstrate robust zero-shot generalization capabilities. By leveraging massive multimodal datasets and cross-modal alignment mechanisms (e.g., CLIP), FIMs seamlessly map textual semantics to visual structures, enabling unprecedented versatility in generation, editing, and understanding.
Models like Sora (Video) and Midjourney v6 have effectively "learned" how light bounces off surfaces and how gravity affects falling objects by simply observing billions of images. 

There are still some significant challenges ahead.
Generative hallucinations, biases embedded in web-scale training data, and the lack of fine-grained control remain active areas of research. Techniques like ControlNet 
and parameter-efficient fine-tuning (PEFT) represent vital steps toward 
more controlled and cost efficient image modeling by FIMs.

Evaluation of image models necessarily involves a trade-off between fidelity and photo-realism. The~quantitative and qualitative evaluation of image processing algorithms have evolved together with the evolution of image models. In the era of handcrafted priors, the sole evaluation metric was Peak Signal-to-Noise Ratio (PSNR). SSIM (Structural Similarity) was proposed to address the well-known issues with PSNR. Since the traditional handcrafted models lacked "naturalness" priors, the results often suffered from visual artifacts such as blur, ringing, blockiness. We transitioned to perceptual metrics such as LPIPS, FID, FVD in the era of generative models. Because of the inherent conflict between fidelity and perceptual realism, known as the perception-distortion trade-off~\cite{bla18}, the bottom line comes to "do~you prefer artifacts or hallucinations?" The general consensus is that the evaluation approach should be task-dependent, i.e., are we doing text/digit recognition (fidelity for trustworthiness) or artistic photography (photo-realism).

\vspace{-10pt}


%

\vspace{-5pt}




%
%
%
%
%
%

\section*{Author}
\vspace{-60pt}
\begin{IEEEbiography}[{\includegraphics[width=1in,height=1.25in,clip,keepaspectratio]{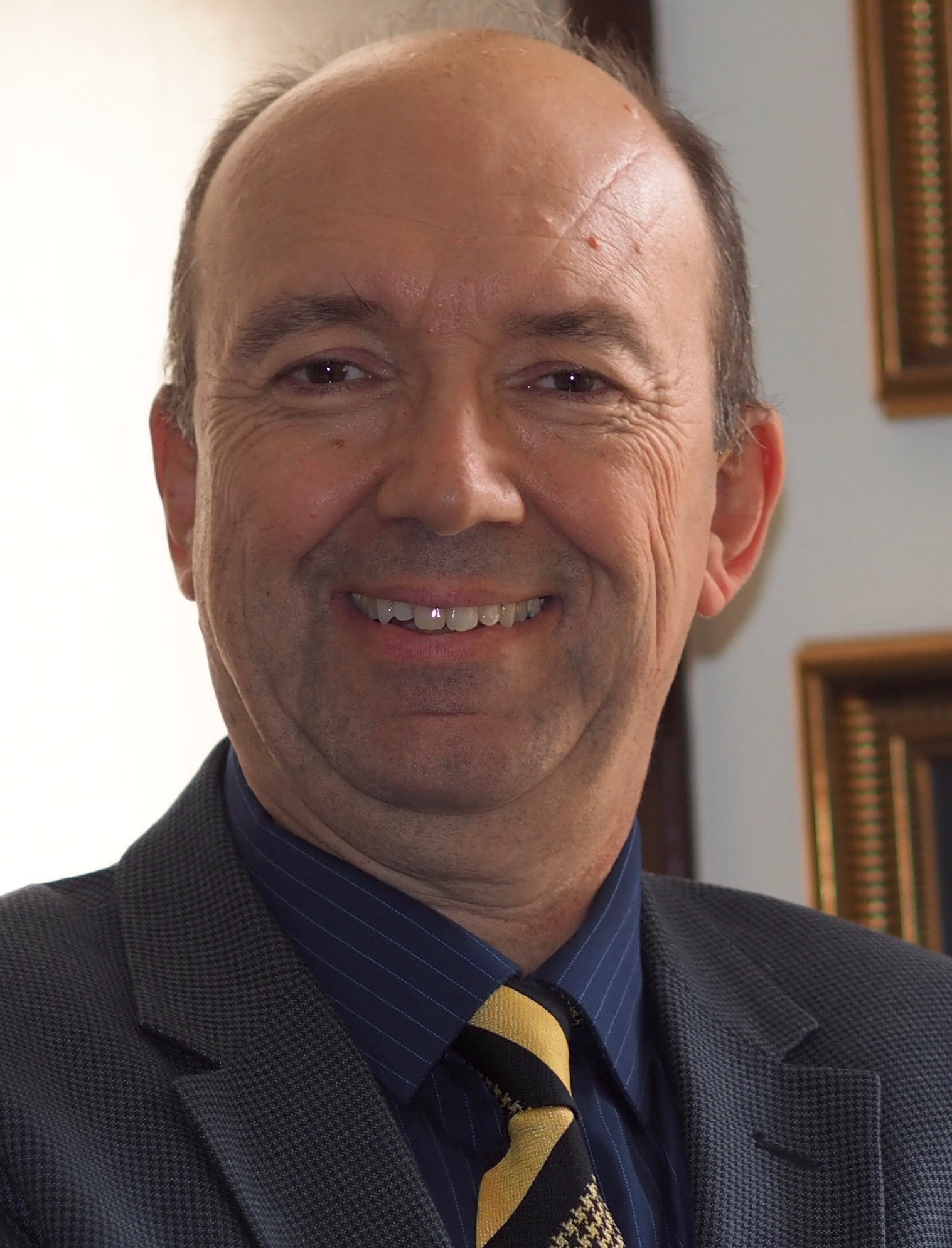}}]{A. Murat Tekalp}(S'80-M'84-SM'91-F'03-LF'24) received Ph.D. degree in Electrical, Computer, and Systems Engineering from Rensselaer Polytechnic Institute (RPI), Troy, New York, in 1984, He was with Eastman Kodak Company, Rochester, New York, from 1984 to 1987, and with the University of Rochester, Rochester, New York, from 1987 to 2005, where he was promoted to Distinguished University Professor. He is currently Professor at Koc University, Istanbul, Turkey. He served as Dean of Engineering between 2010-2013. His research interests are in digital image and video processing, including video compression and streaming, video networking, and deep learning for image/video processing and compression.
\end{IEEEbiography}

\vfill

\end{document}